\documentclass[a4paper,12pt]{article}
\usepackage{graphicx}
\usepackage{yfonts}
\usepackage{mathrsfs}
\usepackage{float}
\usepackage{hyperref}
\usepackage{subcaption}

\usepackage[none]{hyphenat}

\pdfoutput=1
\usepackage{color}
\usepackage{amssymb}
\usepackage{epsfig}
\usepackage{romannum}

\usepackage{graphicx}

\def\L{\mathcal L}
\def\e{\varepsilon}

\newcommand{\wt}{\widetilde}

\usepackage{amsmath}
\usepackage{amsfonts}

\begin{document}

\def\a{\alpha}
\def\b{\beta}
\def\c{\chi}
\def\d{\delta}
\def\e{\epsilon}
\def\f{\phi}
\def\g{\gamma}
\def\h{\eta}
\def\i{\iota}
\def\j{\psi}
\def\k{\kappa}
\def\la{\lambda}
\def\m{\mu}
\def\n{\nu}
\def\o{\omega}
\def\p{\pi}
\def\q{\theta}
\def\r{\rho}
\def\s{\sigma}
\def\t{\tau}
\def\u{\upsilon}
\def\x{\xi}
\def\z{\zeta}
\def\D{\Delta}
\def\F{\Phi}
\def\G{\Gamma}
\def\J{\Psi}
\def\L{\Lambda}
\def\O{\Omega}
\def\P{\Pi}
\def\Q{\Theta}
\def\S{\Sigma}
\def\U{\Upsilon}
\def\X{\Xi}

\def\ve{\varepsilon}
\def\vf{\varphi}
\def\vr{\varrho}
\def\vs{\varsigma}
\def\vq{\vartheta}

\def\dg{\dagger}                                     
\def\ddg{\ddagger}                                   
\def\wt#1{\widetilde{#1}}                    
\def\mt{\widetilde{m}_1}
\def\mti{\widetilde{m}_i}
\def\rt{\widetilde{r}_1}
\def\mtt{\widetilde{m}_2}
\def\mttt{\widetilde{m}_3}
\def\rtt{\widetilde{r}_2}
\def\mb{\overline{m}}
\def\VEV#1{\left\langle #1\right\rangle}        
\def\be{\begin{equation}}
\def\ee{\end{equation}}
\def\ds{\displaystyle}
\def\ra{\rightarrow}

\def\bea{\begin{eqnarray}}
\def\eea{\end{eqnarray}}
\def\NO{\nonumber}
\def\Bar#1{\overline{#1}}


\def\pl#1#2#3{Phys.~Lett.~{\bf B {#1}} ({#2}) #3}
\def\np#1#2#3{Nucl.~Phys.~{\bf B {#1}} ({#2}) #3}
\def\prl#1#2#3{Phys.~Rev.~Lett.~{\bf #1} ({#2}) #3}
\def\pr#1#2#3{Phys.~Rev.~{\bf D {#1}} ({#2}) #3}
\def\zp#1#2#3{Z.~Phys.~{\bf C {#1}} ({#2}) #3}
\def\cqg#1#2#3{Class.~and Quantum Grav.~{\bf {#1}} ({#2}) #3}
\def\cmp#1#2#3{Commun.~Math.~Phys.~{\bf {#1}} ({#2}) #3}
\def\jmp#1#2#3{J.~Math.~Phys.~{\bf {#1}} ({#2}) #3}
\def\ap#1#2#3{Ann.~of Phys.~{\bf {#1}} ({#2}) #3}
\def\prep#1#2#3{Phys.~Rep.~{\bf {#1}C} ({#2}) #3}
\def\ptp#1#2#3{Progr.~Theor.~Phys.~{\bf {#1}} ({#2}) #3}
\def\ijmp#1#2#3{Int.~J.~Mod.~Phys.~{\bf A {#1}} ({#2}) #3}
\def\mpl#1#2#3{Mod.~Phys.~Lett.~{\bf A {#1}} ({#2}) #3}
\def\nc#1#2#3{Nuovo Cim.~{\bf {#1}} ({#2}) #3}
\def\ibid#1#2#3{{\it ibid.}~{\bf {#1}} ({#2}) #3}

\title{
\vspace*{1mm}
{\bf Representing seesaw neutrino models and their motion in lepton flavour space}
\author{
{\Large Pasquale Di Bari$^1$, Michele Re Fiorentin$^2$ and  Rome Samanta$^1$}
\\
$^1$ {\it Physics and Astronomy}, 
{\it University of Southampton,} \\
{\it  Southampton, SO17 1BJ, U.K.} 
\\
$^2$ {\it Center for Sustainable Future Technologies}, \\
{\it Istituto Italiano di Tecnologia, corso Trento 21, 10129 Torino, Italy}
}}
\maketitle \thispagestyle{empty}
\pagenumbering{arabic}

\begin{abstract}
We discuss how seesaw neutrino models can be graphically represented in lepton flavour space.
We examine various popular models and show how this representation  
helps understanding their properties and connection with experimental data
showing in particular how certain texture zero models are ruled out. 
We also introduce a new  matrix, the bridging matrix,  that  brings from the light to the heavy neutrino mass  
flavour basis, showing how this is related to the orthogonal matrix and how different quantities
are easily expressed through it.  We then show how one can
randomly generate  orthogonal and leptonic mixing matrices 
uniformly covering all flavour space in an unbiased way (Haar-distributed matrices).  Using the isomorphism between
the group of complex rotations and the Lorentz group,  we also introduce the concept
of Lorentz boost in flavour space for a seesaw model and how this has 
an insightful physical interpretation.  Finally, as a significant application, we consider $N_2$-leptogenesis. 
Using current experimental  values  of  low energy neutrino parameters, we show that
 the probability that at least one flavoured decay parameter of the lightest right-handed neutrino
is smaller than unity is  about $49\%$ (to be compared with 
the tiny probability that the total decay parameter is smaller than unity, $P(K_{\rm I}< 1)\sim 0.1 \%$,
confirming the crucial role played by flavour effects). On the other hand 
when $m_1 \gtrsim 0.1\,{\rm eV}$ this probability reduces to less than  $5\%$,
showing how also $N_2$-leptogenesis disfavours degenerate light neutrinos. 
\end{abstract}

\newpage
\section{Introduction}

The possibility to identify the origin of neutrino masses and mixing clashes
with  the limited number of low energy neutrino parameters that we can access experimentally, those encoded in the light neutrino mass matrix (three mixing angles, three neutrino masses,
one Dirac phase, two Majorana phases\footnote{The two Majorana phases  are not fully 
measurable. However, a positive signal in neutrinoless double beta decay experiments
would provide an experimental relation between them, placing constraints.}), in comparison with
the large number of theoretical parameters typically introduced by models of new physics. 
Even within  a minimal extension of the Standard Model explaining neutrino masses and mixing, the type I seesaw mechanism \cite{seesaw}, 
there are far too many parameters to obtain definite predictions. This is true unless: 
\begin{itemize}
\item[(i)] either this is embedded within a theoretical framework able to reduce the  number of parameters (top-down approach);
\item[(ii)]  or an explanation of neutrino masses and mixing is linked to other observables (bottom-up approach) such as the matter-antimatter asymmetry of the universe with leptogenesis, parameters in the quark sector (as in grand-unified theories), 
lepton flavour violating processes (within different models), dark matter 
of the universe (from heavy-heavy neutrino mixing or from light-heavy neutrino mixing);
\item[(iii)] or some combination of (i) and (ii) is realised; in this case the top-down and bottom-up 
approaches are complemented and both help to increase the predictive power. 
\end{itemize}
In the case of a pure bottom-up approach one would like to draw model independent conclusions based just
on the experimental information.  From this point of view a useful and widely used tool is the orthogonal parameterisation
of the neutrino Dirac mass matrix within type-I seesaw mechanism since it allows to separate in a unambiguous way the 
light  neutrino parameters, three light neutrino masses and six mixing parameters, from the heavy neutrino parameters
(in the most attractive case of three heavy neutrinos one has three heavy neutrino masses and 
six parameters in the orthogonal matrix). 
Scans within this parameterisation within a particular model or imposing certain constraints
such as successful leptogenesis can lead to interesting bounds 
on low energy neutrino parameters or even to specific predictions. They can also be used to
study the impact of including specific effects in the calculation of the asymmetry or the
validity of certain approximations. 

In this paper we are interested in introducing new general tools for the study and understanding of seesaw models, in particular  how these can be represented in lepton flavour space and randomly generated 
in an unbiased way. 
In Section 2 we show how different models can be graphically represented in flavour space and how this helps
understanding quite easily different properties or aspects of the model, for example whether it can reproduce successfully or not the experimental constraints.  In particular we show how certain models with
textures zeroes are now excluded by the experimental data.  We also review how 
the parameters in the orthogonal matrix relate the light neutrino masses to the heavy neutrino masses
and contain direct information on how fine tuned are the light neutrino masses from the seesaw formula.  
We  introduce a new matrix,  the {\em bridging matrix},  that relates in a simple way the light neutrino mass eigenstates to the lepton states produced by the decays of the  heavy neutrino mass eigenstates. 
In Section 3 we discuss a new parameterisation of the orthogonal matrix 
and of the leptonic mixing matrix such that if no experimental information is imposed, 
a random uniform generation of the parameters produces light and heavy neutrino flavours 
that cover uniformly all lepton flavour space without favouring  any particular flavour direction or region. 
This new parameterisation is based on the isomorphism of the group of complex rotations with the
restricted Lorentz group. In this way we introduce the concept of Lorentz boost in flavour space and, therefore, 
of motion of a model in flavour space with a specified velocity and along a certain direction in flavour space.
This should be meant not as a continuous evolution in flavour space but rather as a property characterising each
flavour model itself. In particular we show that models {\em at rest} in flavour space correspond to models with  
minimal fine-tuning. 

We also apply this new parameterisation to leptogenesis, showing how in this way the distributions of all flavour  decay parameters are identical if no experimental information on the low energy neutrino parameters is imposed
and how these change when current experimental information is imposed. In particular, we consider the  lightest right-handed (RH) neutrino flavoured decay parameters that
play a special role in $N_2$-leptogenesis.  We are able to show, using latest measurements of
neutrino mixing angles,  how the probability that at least one
of the lightest RH neutrino flavoured decay parameters is less than unity is $\sim 49\%$.
 Since this condition determines approximately whether
the asymmetry  produced by the next-to-lightest RH neutrino decays in that flavour can survive the lightest RH
neutrino wash-out, this result shows how successful $N_2$-leptogenesis does not require 
special conditions at all. Finally, in Section 4, we draw the conclusions.

\section{Representing seesaw models in lepton flavour space}

 We consider a traditional extension of the standard model  introducing $N$ right-handed neutrinos 
 $N_{RJ}$ ($J=\Romannum{1},\Romannum{2}, \dots, N$) with Yukawa couplings $h^\nu$ and, allowing for lepton number violation, with Majorana mass matrix $M$.
In the flavour basis where the Majorana mass term 
and the charged lepton Yukawa matrices are both diagonal,
the Yukawa interactions terms for neutrinos and charged leptons plus 
the Majorana mass term  can be written  as 
\be \label{LYM}
- {\cal L}_{Y+M}^{\nu + \ell} =\overline{L_{\a}}\,h^{\ell}_{\a\a}\,{\ell}_{R \a}\, \Phi 
+ \overline{L_{\a}}\,h^{\nu}_{\a J}\,N_{R J}\, \widetilde{\Phi} \,   +
{1\over 2}\,\overline{N_{R J}^{\,c}}\, M_J\, N_{R J} + {\rm h.c.}  \,  ,
\ee
where $L^T \equiv (\nu_L,\a_L)$ are the leptonic doublets, $M_\Romannum{1} \leq \dots \leq M_N$ are the heavy neutrino masses  and we indicated with Greek indexes
the charged lepton flavours, $\a=e,\mu,\tau$, and with Roman indexes the heavy neutrino flavours, $J ={\Romannum{1}}, {\Romannum{2}} \dots, N$. 
After spontaneous symmetry breaking the Higgs vev generates Dirac masses
$m_{D} =v\,h^\nu$ and $m_{\a} =v\,h^{\ell}_{\a\a}$ respectively for neutrinos and charged leptons
so that the total mass term of the Lagrangian for neutrinos and charged leptons can be written as
\be
- {\cal L}^{\ell + \nu}_{m} = \overline{{\a}_{L}}\, m_\a \,{\a}_{R} +  \overline{\nu_{L \a }}\, m_{D \a J} \,  N_{R J} 
+{1\over 2}\,\overline{N_{R J}^{\,c}}\,M_J\, N_{R J} + {\rm h.c.}   \;\; .
\ee
In the  limit $M \gg m_D$, the light neutrino mass matrix is given by the seesaw formula \cite{seesaw}
\be\label{seesaw}
m_{\nu \a\b} = - m_{D \a J} \, M^{-1}_{J} \, m_{D \b J} \,  .
\ee
This is diagonalised by  the (unitary) leptonic mixing matrix $U$
in a way that $m_{\nu \a \b} = - U_{\a i} \, D_{m \, ij} \, U_{\b j}$,
where $D_{m} \equiv {\rm diag}(m_1,m_2,m_3)$. The light neutrino masses $m_1 \leq m_2 \leq m_3$ can then be expressed as
\be
m_i = U^{\star}_{i \a} \, m_{D \a J} \, M^{-1}_{J} \, (m_D^T)_{J \b} \, U^{\star}_{\b i} \,  .
\ee
This expression is equivalent to the orthogonality of the matrix \cite{casasibarra}
\be\label{orthogonal}
\O_{i J} = {(U^\dagger \, m_{D})_{i J} \over \sqrt{m_i \, M_J}} \,  ,
\ee
that provides a useful ({\em orthogonal}) parameterisation of the neutrino Dirac mass matrix 
\be
m_{D \a J} = U_{\a i}\,\sqrt{m_i} \, \O_{i J}\,\sqrt{M_J} \,  .
\ee
The orthogonal matrix elements $\O_{iJ}=|\O_{iJ}|\,e^{i\,{\varphi_{iJ}\over 2}} $ 
have an important physical meaning \cite{masina}. They provide the  fractional contribution to the light neutrino mass $m_i$ from the term proportional to the inverse heavy neutrino mass $M_J^{-1}$ and also, very importantly, they tell how fine-tuned are phase  cancellations in the seesaw formula to get each  $m_i$ as a sum of  terms $ \propto M_J^{-1}$.  
Indeed, it is simple to express each light neutrino mass $m_i$ as \footnote{Notice that 
using the orthogonality of $\O$  one can write
$m_i = m_i \, \sum_J \, \O^2_{iJ}$ and from this and the definition of $\varphi_{iJ}$ one obtains (\ref{miort}).
From Eq.~(\ref{orthogonal}) one can see that $r_{iJ} \propto 1/M_J$.}
\be\label{miort}
m_i =  \overline{m}_i \, \sum_J \, r_{i J} \, e^{i\,\varphi_{iJ}} \,  ,
\ee
where each $r_{iJ} \equiv |\O^2_{iJ}|/\sum_J |\O^2_{iJ}| \propto 1/M_J$ is the fractional contribution
to the neutrino mass  $m_i$ from the heavy neutrino inverse mass $M_J^{-1}$, and
$\overline{m}_i \equiv m_i \, \sum_J |\O^2_{iJ}|$. In this way the quantities
$\gamma_i \equiv \sum_J |\O^2_{iJ}| \geq 1$ can be regarded as a measure of the fine-tuning,
from phase cancellations, that is required to reproduce the light neutrino masses $m_i$. 

If we indicate with  $|L_J \rangle$ the lepton quantum state  produced (at tree level) in the decay of a RH neutrino $N_J$, its charged lepton flavour composition is determined by the neutrino Dirac mass matrix \cite{densitymatrix}
\be
|L_J \rangle = {m_{D \a J} \over \sqrt{(m_D^{\dagger} \, m_D)_{JJ}}} \, |L_{\alpha} \rangle \,  .
\ee
If  we use the leptonic mixing matrix $U$ to express the charged lepton flavour eigenstates in terms of the 
neutrino mass eigenstates, $|L_{\alpha} \rangle = U^{\star}_{\a i} \, |L_{i} \rangle$, we obtain  
\be
|L_J \rangle =  {m_{D \a J} \, U^{\star}_{\a i} \over \sqrt{(m_D^{\dagger} \, m_D)_{JJ}}} \, |L_{i} \rangle 
= {(U^{\dagger}\,m_D)_{i J} \over \sqrt{(m_D^\dagger \, m_D)_{JJ} }} \, |L_{i} \rangle   \,  ,
\ee
showing that the matrix
\be\label{XiJ}
B_{i J} \equiv {(U^\dagger \, m_D)_{i J} \over \sqrt{(m_D^\dagger \, m_D)_{JJ}}}   
\ee
operates the transformation between the lepton flavour basis  determined by the neutrino mass eigenstates  to 
that one determined by heavy neutrino lepton flavour states.\footnote{If one considers the  lepton doublet  fields, rather than the states, one has $L_J = B^\star_{J i}\, L_i$. Notice also that the matrix $U^{\dagger}\,m_D$ is the Dirac neutrino mass matrix in a flavour basis where both light neutrino and heavy neutrino mass matrices are diagonal \cite{fujii}.
The $B$ matrix is obtained properly normalising $U^{\dagger}\,m_D$ and it basically 
bridges the energy gap between low and high energy states,
more precisely bringing from low to high energy states. For this reason it could be also referred to as 
the {\em beanstalk matrix}, from the beanstalk narrated in the {\em Story of Jack and the Beanstalk}.}
In terms of the orthogonal matrix one finds easily
\be
B_{i J} = \sqrt{m_i \over \widetilde{m}_J}\,\O_{i J} = {\sqrt{m_i}\,\O_{iJ} \over \sqrt{\sum_k \, m_k \,|\O_{kJ}|^2} }\,  ,
\ee
where we introduced the effective neutrino masses \cite{plumacher,fujii}
\be\label{effectivenmasses}
\widetilde{m}_J \equiv {(m_D^{\dagger} \, m_D)_{JJ} \over M_J} = \sum_k \, m_k \, |\O_{kJ}|^2 \,  .
\ee
This shows that for $N=3$ the matrix $B$ contains nine parameters: the three light neutrino masses and the six parameters  in the orthogonal matrix.
These are indeed the $3\times 3$ parameters necessary to determine
the flavour compositions of the heavy neutrino flavour states $|L_J \rangle$. 

The probability that  a lepton $L_J$ is  measured as a lepton $L_{I \neq J}$ or, equivalently, the interference probability
between a heavy neutrino $N_J$ and a heavy neutrino $N_{I \neq J}$, can be simply expressed (at tree level) in terms of $B$ as
\be\label{pIJ}
p^0_{IJ} \equiv |\langle L_J | L_I \rangle|^2 = \left|\sum_k \,  B^\star_{k J}\, B_{k I}\right|^2   \,   ,
\ee
and one can immediately verify from Eq.~(\ref{XiJ}) that $p^0_{JJ}=1$. The nought in the upper script  
indicates that they are calculated at tree level. 
On the other hand the probability that a lepton $L_I$ is measured in a 
charged lepton flavour $\a =e,\mu,\tau$ is given by
\be\label{p0Ial}
p^0_{J\a} \equiv |\langle L_\a | L_J \rangle|^2 
={|m_{D \a J}|^2 \over (m_D^{\dagger} \, m_D)_{JJ}} =  
\left| \sum_k  U_{\a k} \, B_{k J} \right|^2 =
{|\sum_k \, \sqrt{m_k} \, U_{\a k} \, \O_{k J}|^2 \over \sum_k \, m_k \, |\O_{k J}|^2} \,  .
\ee
These expressions for the probabilities clearly show the physical meaning of the $B$ matrix
as a transformation matrix between the light and the heavy neutrino flavour basis. 
    
The seesaw formula is invariant under a generic unitary flavour transformation of the LH fields 
$\nu_{L \a'} = V_{L\a'\a} \, \nu_{L \a}$, so that one can write in the new flavour basis
\be
m'_{\nu} = - m'_D \, {1 \over D_M} \, m_D^{'T}  \,   ,
\ee
where $D_M \equiv {\rm diag}({M_\Romannum{1},M_\Romannum{2},\dots,M_N})$ and
$m'_{D \a' J} = V_{L \, \a' \a}  \, m_{D\a J}$, while the transformed light neutrino mass matrix is given by 
$m'_{\nu\,\a'\b'} = V_{L \, \a'\a} \, m_{\nu\,\a\b} \, (V_L^{T})_{\b\b'}$. 
In this new basis the charged lepton mass matrix is in general non-diagonal. 
The orthogonal matrix $\O$ and the bridging matrix $B$ are of course invariant under this 
change of lepton flavour basis, since they are by definition transformations between 
the light and the heavy neutrino  flavour bases and, therefore, are independent 
of which lepton flavour basis is chosen to represent the lepton fields and neutrino Dirac mass matrix. 
Therefore, in terms of the transformed Dirac mass matrix, they can be simply written  as 
\be\label{orthogonalbridgingnew}
\O_{i J}  =  {(W^{\dagger}\, m'_{D})_{i J} \over \sqrt{m_i \, M_J}} \,  , \;\;\;\;\;\;
B_{i J}   =  {(W^{\dagger}\, m'_D)_{i J} \over \sqrt{(m_D^{'\dagger} \, m'_D)_{JJ}}}    \,  ,
\ee
where we introduced the unitary matrix $W_{\a' i} \equiv V_{\a'\a}\,U_{\a i}$ that brings from the
light neutrino mass basis to the new generic primed flavour basis. 

{\em Neutrino Yukawa basis}. A particularly important example of lepton flavour basis, useful especially to describe a model,
is represented by the neutrino Yukawa basis. This is the basis where the neutrino Dirac mass matrix
is diagonal. In general the change to this basis has to be done transforming simultaneously
both the LH neutrino fields and the RH neutrino fields by means of a bi-unitary transformation, 
$\nu_{L \ell} = V^Y_{L \ell \a} \, \nu_{L \a}$ and $N^Y_{R \ell} = U^Y_{R \ell I}\, N_{R I}$ respectively
($\ell=a,b,c$),  in a way that
\be\label{yukawa}
m_{D} = V_L^{Y \dagger} \, D_{m_D} \, U^Y_R \,  ,
\ee
where $D_{m_D} \equiv {\rm diag}(m_{Da},m_{Db},m_{Dc})$ and  $m_{Da} \leq m_{Db} \leq m_{Dc}$ 
are the Dirac masses.  The Yukawa basis
has important physical properties. First of all whether the leptonic mixing matrix is generated either
in the LH sector or in the RH sector is clearly something encoded by $V_L^Y$
and $U_R^Y$ respectively. If there is no right-right Majorana mass term (the Dirac neutrino case) 
then the light neutrino masses would be simply given by the Dirac masses (i.e., $m_1=m_a$, $m_2=m_b$, $m_3 = m_c$)
and the leptonic  mixing matrix would be simply given by $U=V_L^{Y\dagger}$: the Yukawa basis
would simply coincide with the light neutrino mass basis. 
This would be still true when the Majorana mass term is turned on, the case of our interest,
and $U_R^Y=I$, corresponding to say that the Majorana and the Dirac mass matrices are diagonalised in the same
basis. The only difference would be that in this case one has seesawed neutrino masses  $m_i = m^2_{D \ell}/M_I$
with $\ell =a,b,c$. 
Vice-versa, if $V_L^Y = I$, then leptonic mixing
can only stem by a $U_R^Y \neq I$, as it can be immediately understood from the see-saw formula.
Another important physical property of the Yukawa basis is that it sets
the right basis where to describe medium effects in the description of 
RH-RH neutrino mixing in the early universe, proposed for example to be either the source of baryogenesis 
in the ARS mechanism \cite{ARS} or of dark matter-genesis in \cite{anisimov}, since
the effective potential due to medium effects are diagonal in the Yukawa basis. 
Therefore, the RH neutrino mixing matrix should be identified with  $U_R^Y$, at least  
in the absence of other (non-standard) RH neutrino interactions. 
Finally, notice that the neutrino Yukawa basis provides clearly the reference basis to compare the neutrino
Yukawa interactions with those of other massive fermions and in case impose certain relations as
in $SO(10)$-inspired models \cite{SO10inspired} where the neutrino Dirac mass matrix is `not too different' from 
the up quark Dirac mass matrix. Also, as we will see, often this is the right basis where to impose certain conditions rising from symmetries of the model, such as textures zeros or other relations on the mass matrices
of other fermions.

Let us now consider a few interesting examples of lepton flavour bases associated to specific 
classes of models. These are graphically shown in the panels of Fig.~1 where we used the light neutrino flavour 
basis as a reference frame.

{\em Charged lepton flavour basis}. In panel (a) we show the usual charged lepton flavour basis and  how this can be 
obtained, modulo the three phases, from the light neutrino flavour basis by means of three Euler rotations 
defining the  three mixing angles in the leptonic mixing matrix \cite{kingreview}.

{\em Generic heavy neutrino flavour basis}. In panel (b) we show a generic heavy lepton flavour basis  
that, in general, is not orthonormal. We have 
defined the angles $\theta_{IJ}$ simply in such a way that $p^0_{IJ} = \cos \theta_{IJ}$. 
\begin{figure}
\begin{center}
\begin{subfigure}[t]{0.39\textwidth}
\includegraphics[width=\textwidth]{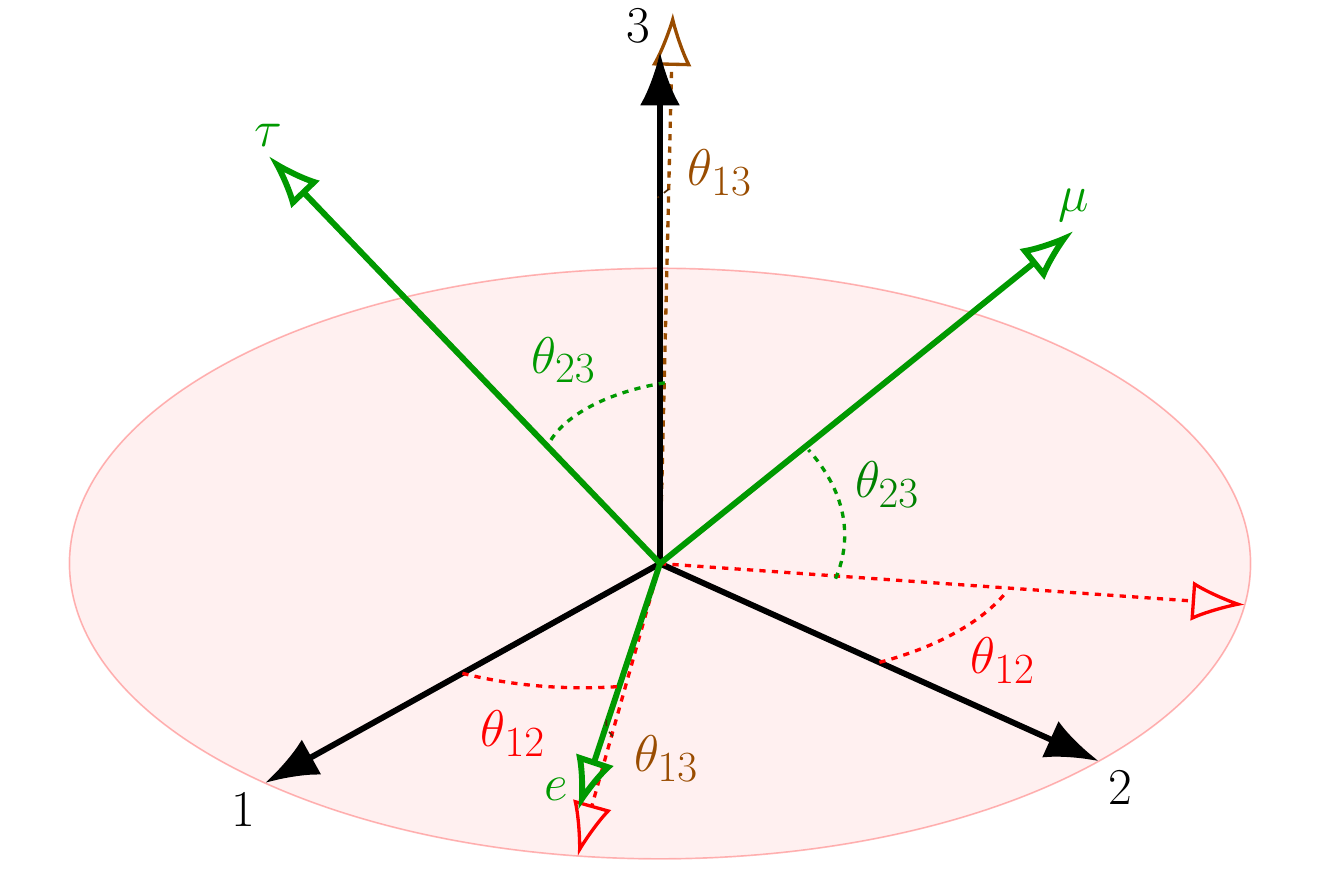} 
\subcaption{Charged lepton flavour basis and mixing angles.}
\end{subfigure}
\hspace*{10mm}
\begin{subfigure}[t]{0.39\textwidth}
\includegraphics[width=\textwidth]{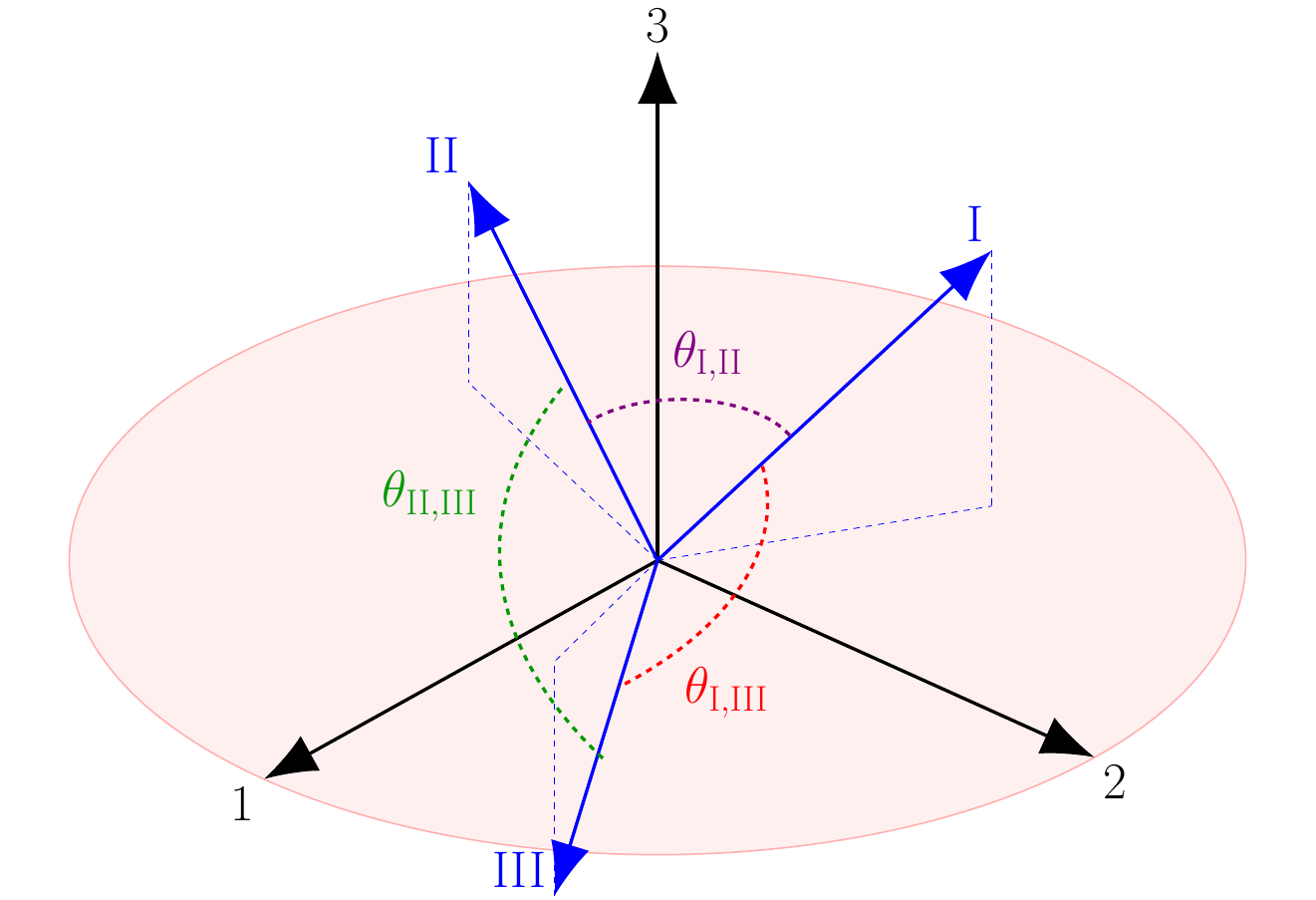} 
\subcaption{Generic lepton heavy neutrino flavour basis.}
\end{subfigure}
 \\ \vspace*{10mm}
\begin{subfigure}[t]{0.39\textwidth}
\includegraphics[width=\textwidth]{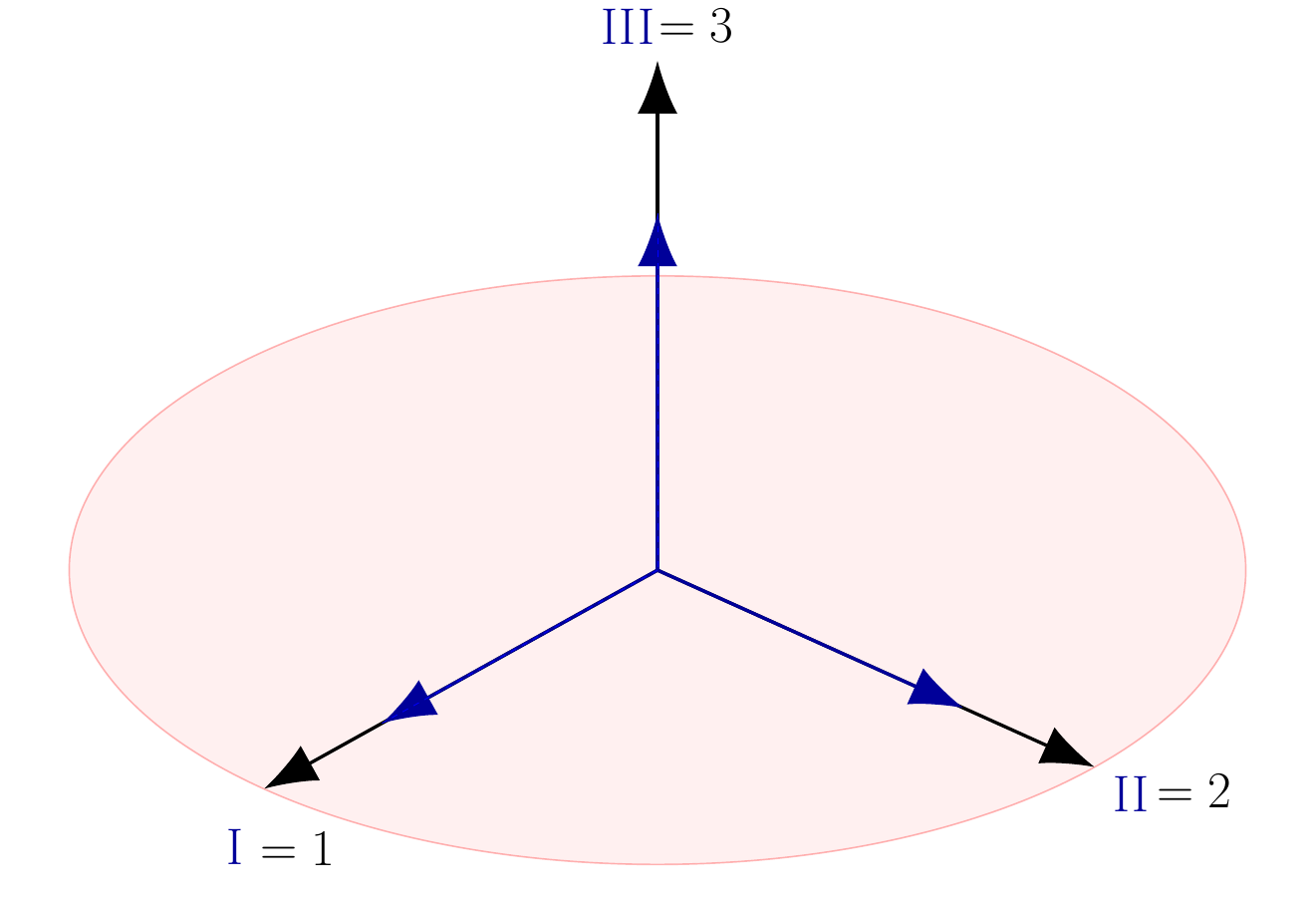} 
\subcaption{Orthonormal heavy neutrino flavour basis 
(necessarily) coinciding with the light neutrino flavour basis.}
\end{subfigure}
\hspace*{10mm}
\begin{subfigure}[t]{0.39\textwidth}
\includegraphics[width=\textwidth]{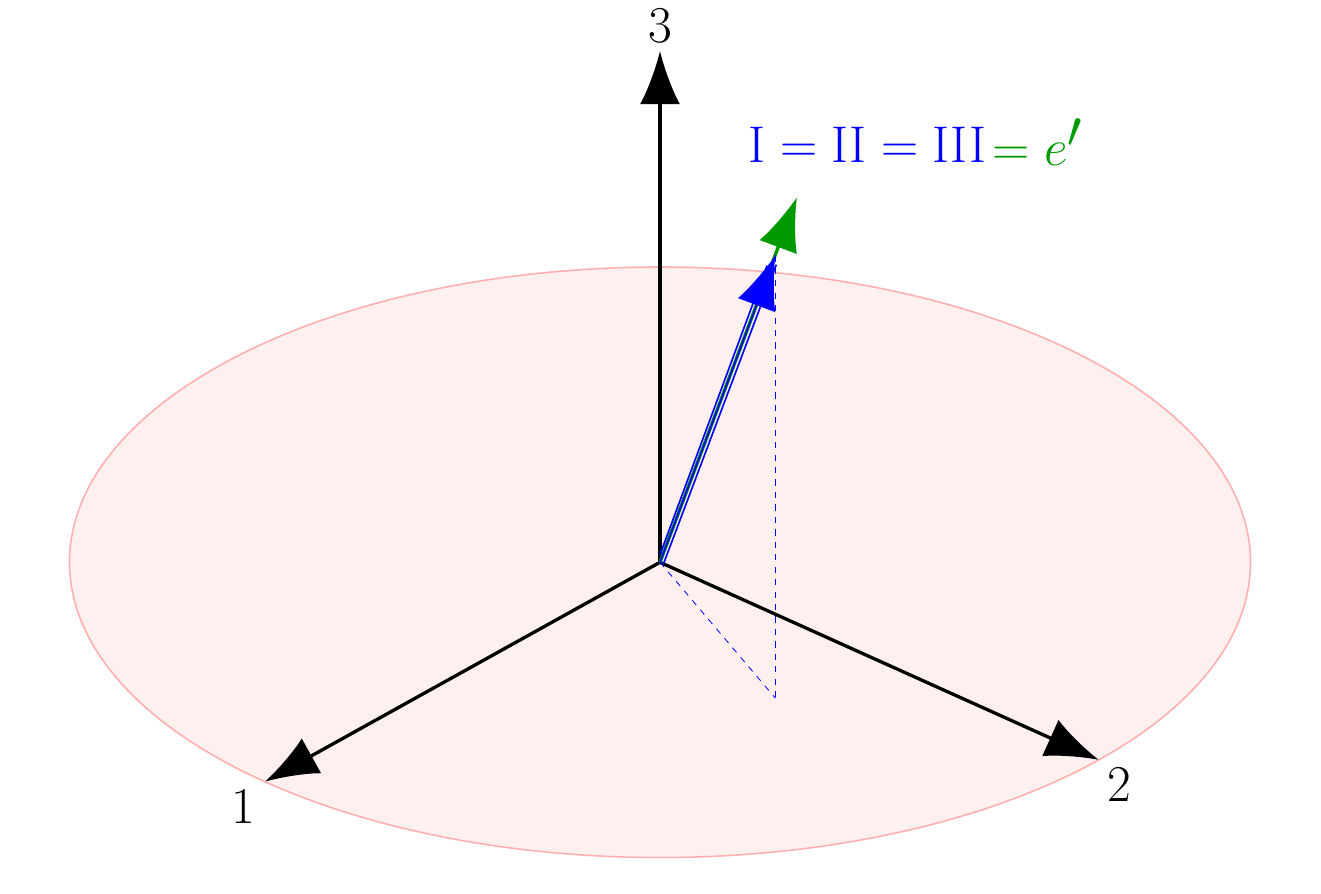} 
\subcaption{Example of three coinciding lepton heavy neutrino flavours.}
\end{subfigure}
 \\ \vspace*{10mm}
\begin{subfigure}[t]{0.39\textwidth}
\includegraphics[width=\textwidth]{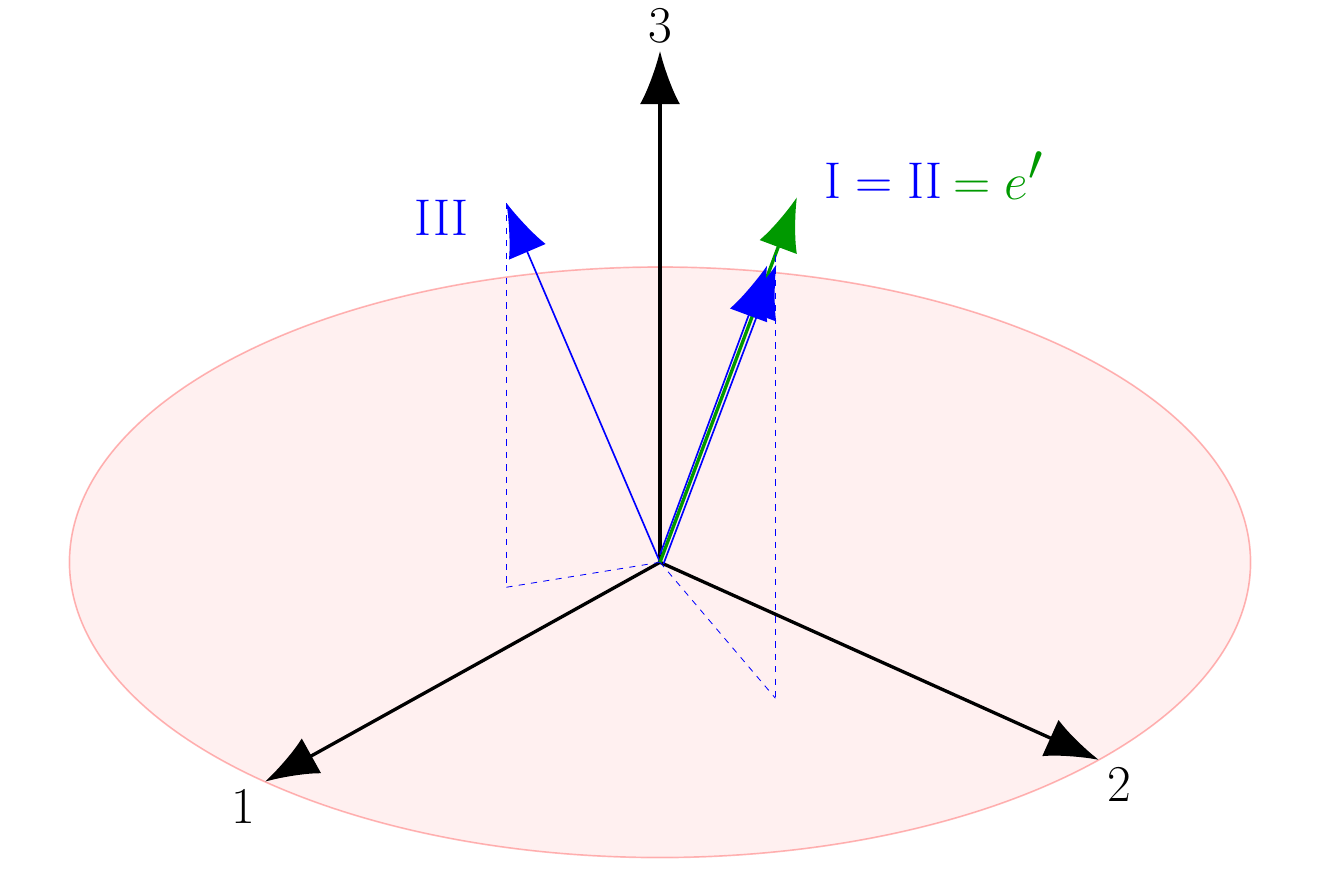} 
\subcaption{Example of two coinciding lepton heavy neutrino flavours.}
\end{subfigure}
\hspace*{10mm}
\begin{subfigure}[t]{0.39\textwidth}
\includegraphics[width=\textwidth]{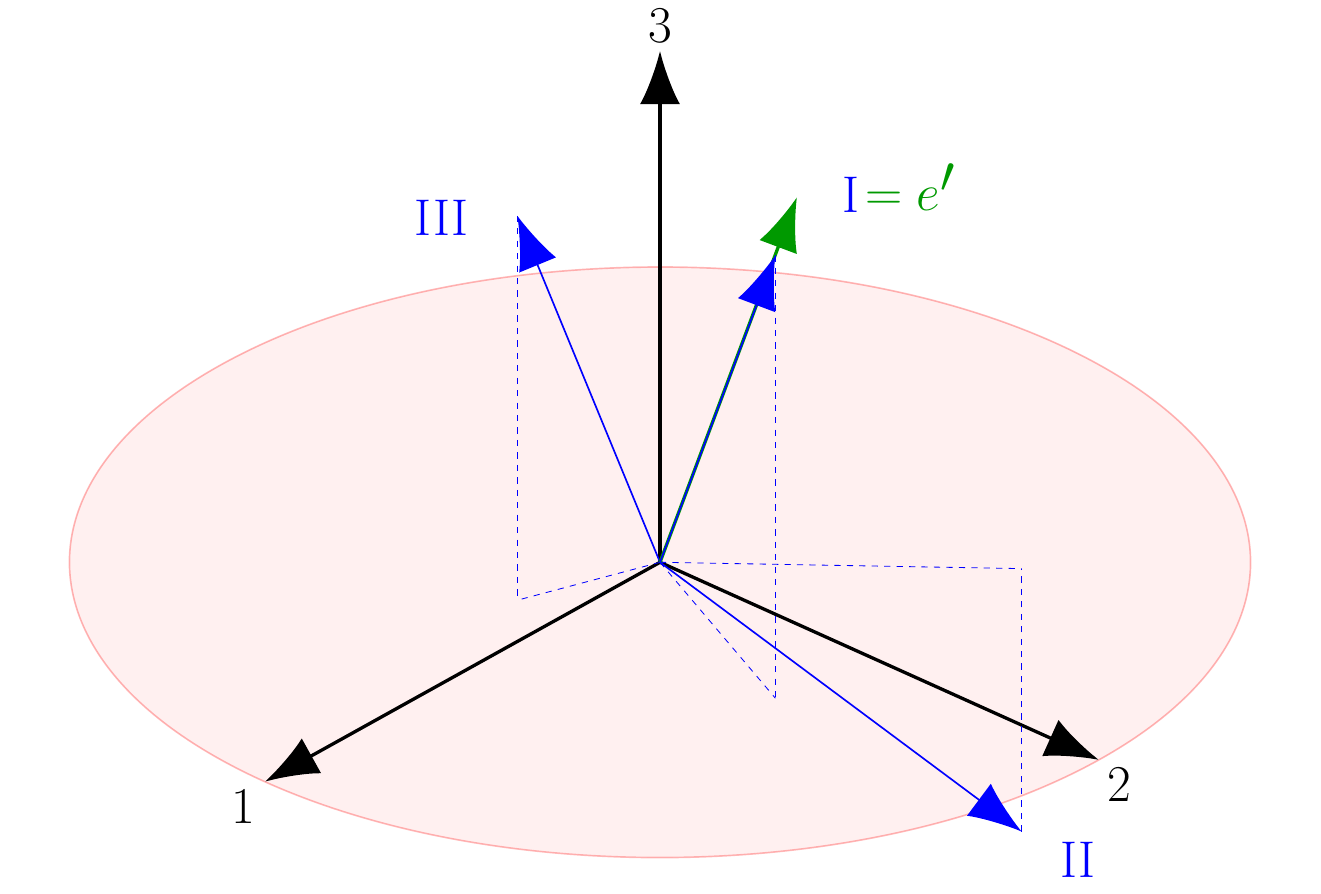} 
\subcaption{Lepton flavour basis where one flavour $e'$ is made coinciding with one of the
heavy neutrino flavours.}
\end{subfigure}
 \caption{Examples of lepton flavour bases with the light neutrino flavour basis as reference basis.} 
 \label{Upmns}
\end{center}
\end{figure}
If the heavy neutrino flavour basis is orthonormal, then
$p^0_{IJ}= \d_{IJ}$ and the equation (\ref{pIJ}) 
correctly shows that in this case $B$ is unitary (however, we  
show in the following that in this case it has necessarily to coincide with the identity or any permutation matrix).  

{\em Coincident light and heavy neutrino flavour bases}. A special case, shown graphically in 
panel (c), corresponds to have $B=P$  (where $P$ here and elsewhere is the permutation matrix), 
so that the heavy neutrino flavour basis 
basically coincides with the light neutrino flavour basis.
In this case one can easily see that necessarily also  both $\O=P$  and  $U_R^Y = P$.
This corresponds to the situation described before 
when the neutrino masses are given simply by $m_i = m^2_{D\ell}/M_J$ and $U=V^{Y \dagger}_L$.
Therefore, in this case necessarily the Yukawa basis also coincides (modulo axes permutations) with the light
and heavy neutrino flavour bases. Indeed, it is correct to say that, since the heavy neutrino flavour basis is aligned with the Yukawa basis, then the resulting light neutrino basis, from the seesaw formula, is also coinciding.
This situation corresponds to what has been called {\em limit of exact dominance}  in \cite{geometry}
or {\em form dominance}  in \cite{muchunking}.
In this case heavy neutrinos do not mix and do not interfere in decays and indeed all $C\!P$
asymmetries, both total \cite{geometry} and flavoured \cite{muchunking}, vanish. 
For this reason some departure from form dominance is necessary if one wants to
realise leptogenesis.  This class of models typically emerges when a non-Abelian flavour symmetry 
is imposed  \cite{bdfn}
in a way that $D_N(g) \, m^{\dagger}_D \, m_D \, D_N(g) = m^{\dagger}_D \, m_D$, where $D_N(g)$ is 3-dim
irreducible representation of the non-Abelian flavour symmetry group $G$ acting on the RH neutrinos 
and $g$ is a generic group element.
In this case the first Shur's lemma implies  $m_D^\dagger \,  m_D= \la^2_D \, P$, where $P$ is the permutation matrix  and 
$\la_D$ is the value of the degenerate Dirac neutrino masses, in a way that 
$m_i = \lambda _D^2 / M_J$ realising form dominance 
corresponding indeed to $\O=P$.  
From Eq.~(\ref{orthogonal}) one can see that the fact  that  $V_L^Y = U^{\dagger}$ 
is consistent with having $\Omega = P$ (and from Eq.~(\ref{XiJ}) that $B = P$).
In order to have successful leptogenesis the flavour symmetry
has to be broken and the $C\!P$ asymmetries are related to the symmetry breaking parameter \cite{jenkins,bdfn}.

One can wonder whether there  can be models, generalising $\O= B =P$, 
characterised by a generic orthonormal  heavy neutrino flavour basis
that does not coincide with the light neutrino flavour basis. However, it is easy to show that this is impossible.
The reason is that if the heavy neutrino flavour basis is orthonormal, then this has necessarily to coincide
with the Yukawa basis since one can always find a matrix $V_L$ that brings to a basis where 
$m'_D$ is diagonal and, therefore, this necessarily implies $U_R = P$. However, in this case from the seesaw formula one immediately finds  $V_L = U^\dagger_L$ and, therefore, the heavy neutrino
flavour basis has necessarily to coincide with the light neutrino flavour basis, as confirmed also by the fact that
one has $\O = B= P$.
%

{\em Three coinciding heavy neutrino flavours}.
An opposite limit case, shown in panel (d), is realised when all three lepton heavy neutrino 
flavours coincide,  i.e. ${\Romannum{1}}= {\Romannum{2}}={\Romannum{3}}=e'$,
meaning that all three heavy neutrinos decay into leptons with the 
same flavour  $e'$. It is  easy to prove  that this case is excluded by the experimental data since
one can always perform a transformation, operated by a unitary matrix $V'_{L }$ acting on the lepton doublets, 
from the (charged lepton) flavour basis $(e,\mu,\tau)$
to a new orthonormal flavour basis $(e',\mu',\tau')$ where $e'$ coincides then with the common heavy neutrino flavour. 
In this new flavour basis the neutrino Dirac mass matrix takes the very simple form
\be\label{mD1flavour}
m_D' = V'_L \, m_D =  \left( \begin{array}{ccc}
m_{De' {\Romannum{1}}} & m_{De' \Romannum{2}}  & m_{De' {\Romannum{3}}}   \\
0 & 0 &  0\\
0 & 0 &  0
\end{array}\right)
 \,   ,
\ee
where $V'_L$ is a unitary matrix that transforms the 
lepton doublets from the charged lepton flavour basis to the new flavour basis.
From the seesaw formula one can see immediately that this form implies $m_1 = m_2 =0$,\footnote{This is something expected since the matrix (\ref{mD1flavour}) has rank 1.} and therefore this case  is excluded since it cannot reproduce both solar and atmospheric  neutrino mass scales. 

{\em Two coinciding heavy neutrino flavours}. We can now consider a less special case where only two lepton heavy neutrino flavours coincide, 
while the third does not and is generic. For example,  we can consider 
${\Romannum{1}}={\Romannum{2}}$.  
In this case we can always find a transformation, still operated by a unitary matrix $V_L'$, from  $(e,\mu,\tau)$ 
to a new orthonormal flavour basis  $(e',\mu',\tau')$ where $e'={\Romannum{1}}={\Romannum{2}}$.  
This case is shown graphically in   panel (e) of Fig.~1. 
In this new flavour basis the Dirac mass matrix takes the form
\be\label{mD2flavour}
m_D' = V'_L \, m_D =  \left( \begin{array}{ccc}
m_{De' {\Romannum{1}}} & m_{De' \Romannum{2}}  & m_{De' {\Romannum{3}}}   \\
0 & 0 & m_{D \mu'  {\Romannum{3}}}  \\
0 & 0 & m_{D \tau'  {\Romannum{3}}} 
\end{array}\right)
 \,   .
 \ee
This form for $m'_D$ can successfully reproduce all low energy neutrino data for a generic $e'$. 
However, if the flavour $e'$ coincides with one of the charged lepton flavours 
(in this case $V'_L = P$), then the number of parameters gets considerably reduced and  
one has to verify for each case, whether it is possible to reproduce the low energy neutrino data. 
For example, if $e'=e$, then one obtains a seesaw model that implies a light neutrino mass matrix of the form respecting the so-called {\em strong scaling ansatz} \cite{moharode,rome},
leading necessarily to a vanishing $\theta_{13}$ now excluded by the data.  
This is only one out of nine cases corresponding to have $|L_I \rangle=|L_J \rangle = |L_\alpha\rangle$ with
$I \neq J$ and $\a =e,\mu,\tau$.  By inspection we have checked 
that also all the other eight cases, listed explicitly in Appendix A, are excluded, since
they give rise to a light neutrino mass matrix that is either again respecting the scaling ansatz made in \cite{moharode} or has some similar scaling property also leading to unacceptable low energy neutrino data 
(see Appendix A for more details).

A popular class of seesaw models where the number of parameters is considerably reduced 
is the two right-handed neutrino limit \cite{2RHnu}.  This can be obtained from the three RH neutrino case  
either in the limit of very large  heaviest RH neutrino mass $M_3 \gg 10^{15}\,{\rm GeV}$ 
or if one of the three RH neutrinos has negligible Yukawa couplings.
In both cases one has $m_1 \ra 0$ and effectively the heaviest RH neutrino decouples from the 
seesaw formula. In this case  one effectively  obtains a  two RH neutrino formula with a $3 \times 2 $ Dirac neutrino mass matrix.   In this case the number of seesaw parameters
reduces from eighteen to eleven. These are still too many to lead to predictions on the mixing parameters
and usually one has to add some additional condition to this extent.  
For example, one could again consider a situation when both the two  heavy neutrino flavours are aligned. However, analogously to the three RH neutrino case where all three heavy neutrino flavours are aligned, one would get a second vanishing light neutrino, so that one cannot reproduce both the solar and the atmospheric neutrino mass scales. Within these two RH neutrino models one can further reduce the number of parameters  again imposing texture zeros in the neutrino Dirac mass matrix $m_D$, i.e., in the charged lepton flavour basis.
In this case it has been shown that models with more than two textures zeros are all ruled out by the data and even among all possible models with two texture zeroes  only one  is still marginally allowed since it requires inverted hierarchy, now disfavoured at approximately $3\sigma$ \cite{nufit}, while all possibilities leading to normal hierarchical neutrino masses do not reproduce the measured values of the mixing angles 
\cite{occam,barreiros}.

{\em Lepton flavour basis leading to two texture zeros in the Dirac mass matrix}. Finally, let us conclude saying that of course one can always find a flavour basis $(e',\mu',\tau')$ where $m'_D$ has two textures zero, since one can always align one flavour along one of the heavy neutrino flavours, for example in a way that $e' = I$ as represented in panel (f) of Fig.~1. 

\section{Motion in lepton flavour space}

The orthogonal parameterisation (see Eq.~(\ref{orthogonal})) is a useful tool that allows
to scan over the (unknown) parameters in the orthogonal matrix and the RH neutrino masses 
taking into account the experimental information from low energy neutrino experiments
also in combination with other phenomenological conditions (e.g., successful leptogenesis, 
reproducing the observed dark matter abundance, respecting constraints
on rates of lepton flavour violating processes). 

The scans are traditionally done using a parameterisation of the leptonic mixing matrix
in terms of three Euler rotations  (two real ones and one complex), defining the three mixing 
angles $\theta_{ij}$, the $C\!P$ violating Dirac phase $\d$ and two $C\!P$ violating 
Majorana phases $\rho$ and $\sigma$, explicitly \cite{kingreview}
\be
U = \left( \begin{array}{ccc}
1 &  0 & 0 \\
0 & c_{23} & s_{23} \\
0 & - s_{23}  & c_{23}
\end{array}\right) \,  
\left( \begin{array}{ccc}
c_{13} & 0  & s_{13}\,e^{-{\rm i}\,\d} \\
0 & 1 & 0 \\
- s_{13}\,e^{{\rm i}\,\d} & 0 & c_{13}
\end{array}\right) \,  
\left( \begin{array}{ccc}
c_{12} &  s_{12} & 0 \\
-s_{12} & c_{12} & 0 \\
0 & 0 & 1
\end{array}\right) \,  
\left( \begin{array}{ccc}
e^{i \rho} &  0 & 0 \\
0 & 1 & 0 \\
0 & 0 & e^{i \sigma}
\end{array}\right) 
\ee
where $s_{ij} \equiv \sin \theta_{ij}$ and $c_{ij} \equiv \cos\theta_{ij}$.
The orthogonal matrix is usually analogously parameterised  as the product
of three complex rotations,
\be\label{Omega}
\O =
\zeta \, 
\left( \begin{array}{ccc}
1 &  0 & 0 \\
0 & \cos z_{23} & \sin z_{23} \\
0 & - \sin z_{23}  & \cos z_{23}
\end{array}\right) \,  
\left( \begin{array}{ccc}
\cos z_{13} & 0  & \sin z_{13} \\
0 & 1 & 0 \\
- \sin z_{13} & 0 & \cos z_{13}
\end{array}\right) \,  
\left( \begin{array}{ccc}
\cos z_{12} & \sin z_{12} & 0 \\
-\sin z_{12} & \cos z_{12} & 0 \\
0 & 0 & 1
\end{array}\right)  \,  ,
\ee
where the $z_{ij}$'s are three complex mixing angles and the overall sign, $\zeta = \pm 1$,
takes into account  two possible different options (branches), one with
positive determinant and one with negative determinant. The three complex
mixing angles can in turn be parameterised in terms of their real
and imaginary parts writing $z_{ij} = x_{ij} + i \, y_{ij}$.

Points in flavour space, for a fixed set of light and heavy neutrino masses, can then
be  obtained generating random uniformly the values of the three mixing angles, the three phases,
the three real and imaginary parts of the complex angles.  The result of this  random generation of points in flavour space is
shown  in the top-left panel of Fig.~2 for arbitrary values of the mixing angles and of the low energy phases, i.e., 
without imposing any experimental constraint.  More precisely we show, with triangular plots, the probability density distribution 
in the space of the charged lepton flavour probabilities $p^0_{{I}\a}$ ($\a=e,\mu,\tau$) for the lightest RH neutrino.
As one can see, though we randomly uniformly generated the values of the parameters, the distribution
exhibits a strong inhomogeneity toward large values of $p^0_{{I}e}$.  This is of course an unpleasant 
feature if one wants to get unbiased flavour distributions where the flavour dependence originates 
only from the experimental data and/or from the properties of a model and is not an artefact of 
how the random generation of points is performed.
\begin{figure}
\begin{center}
\hspace*{0mm}\includegraphics[scale=0.26]{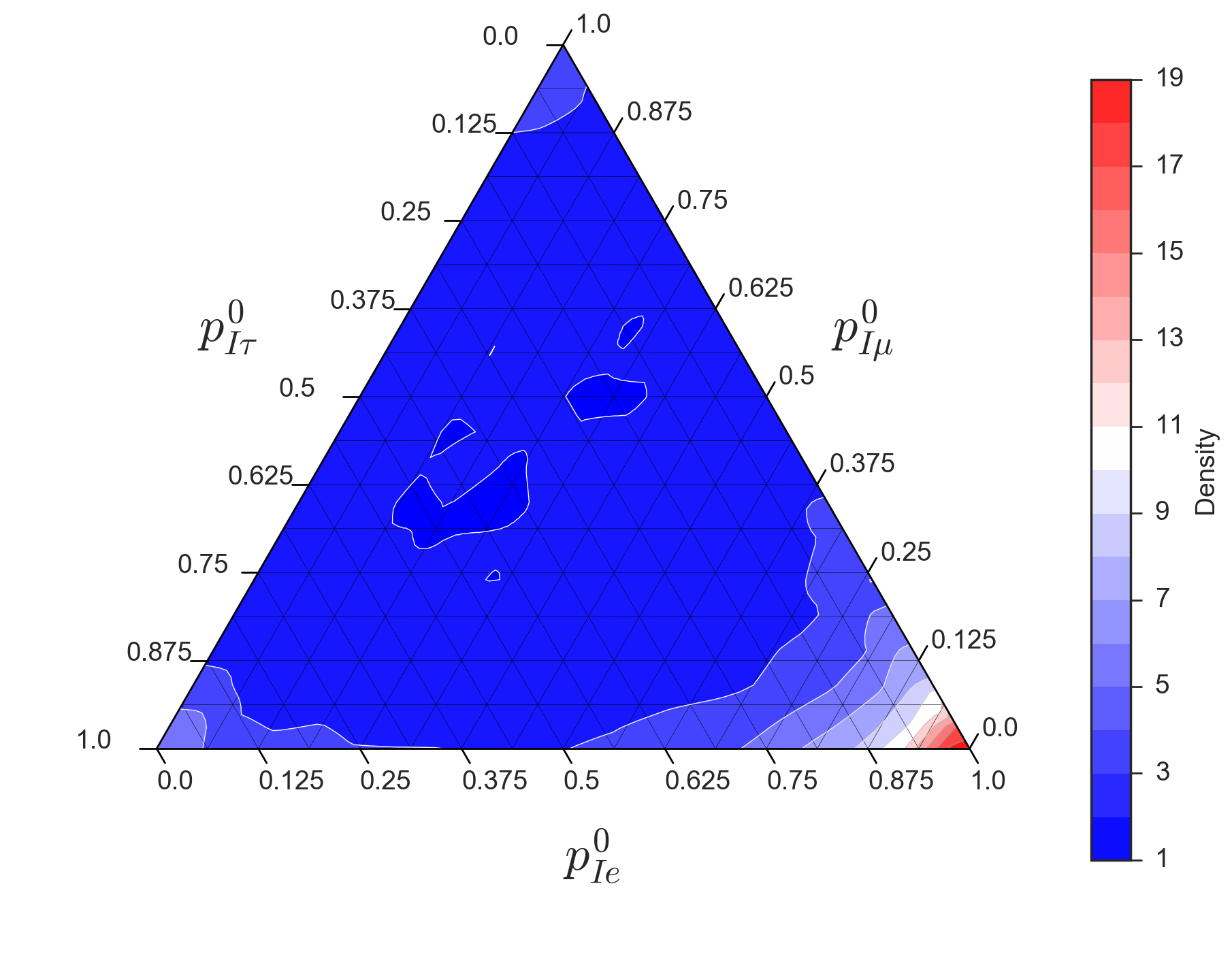} \hspace*{2mm}
\includegraphics[scale=0.26]{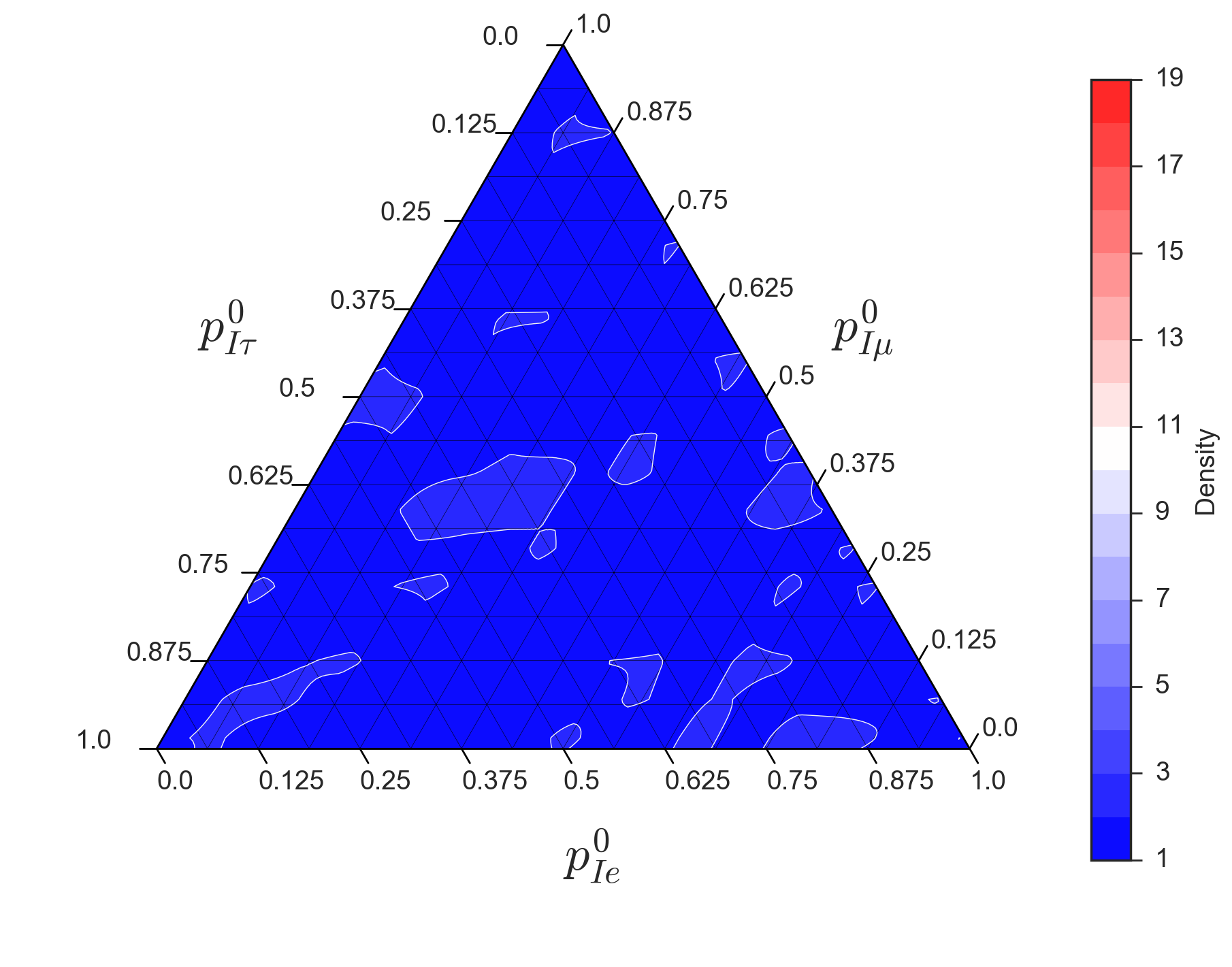} \\
\hspace*{0mm}\includegraphics[scale=0.26]{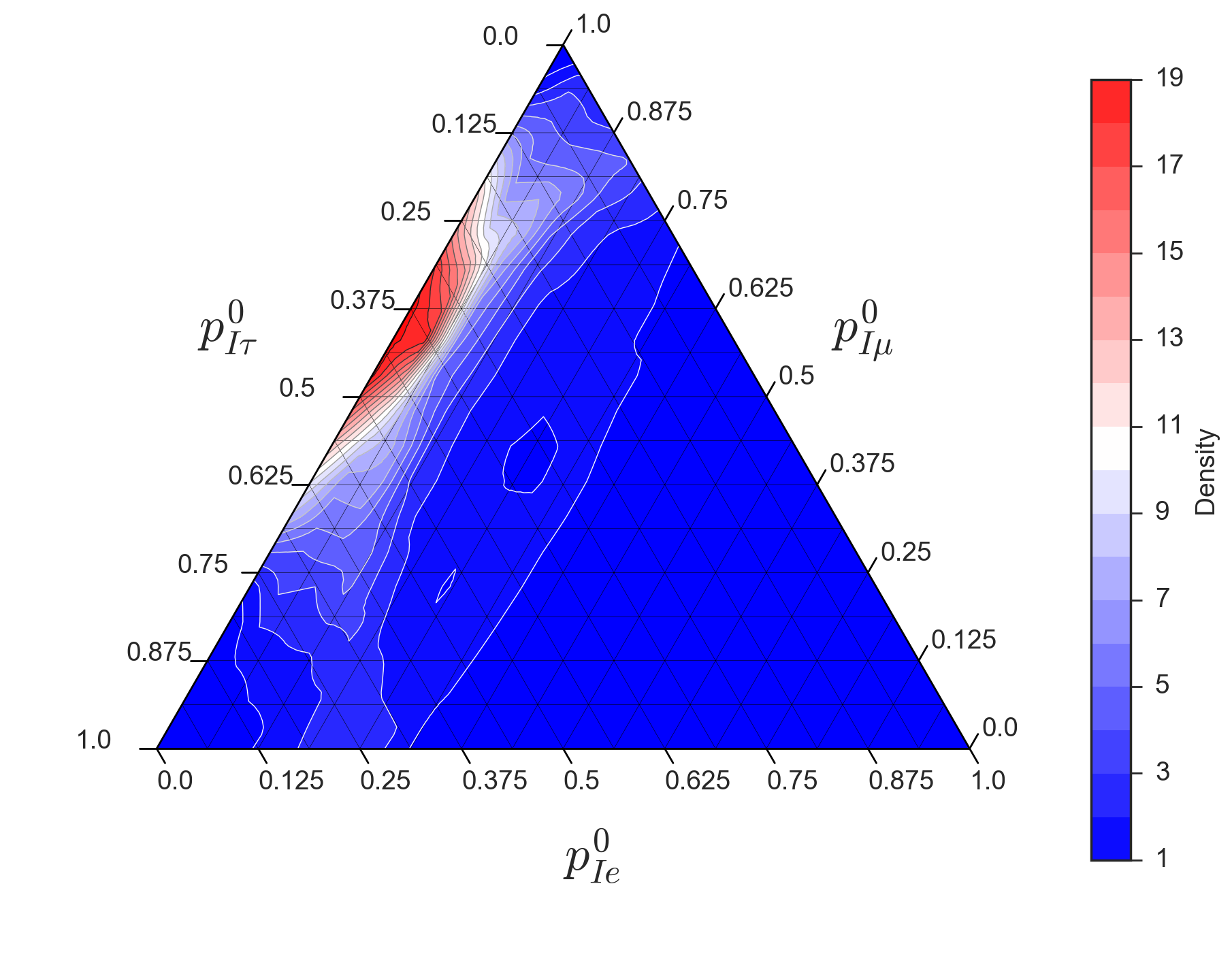} \hspace*{2mm}
\includegraphics[scale=0.26]{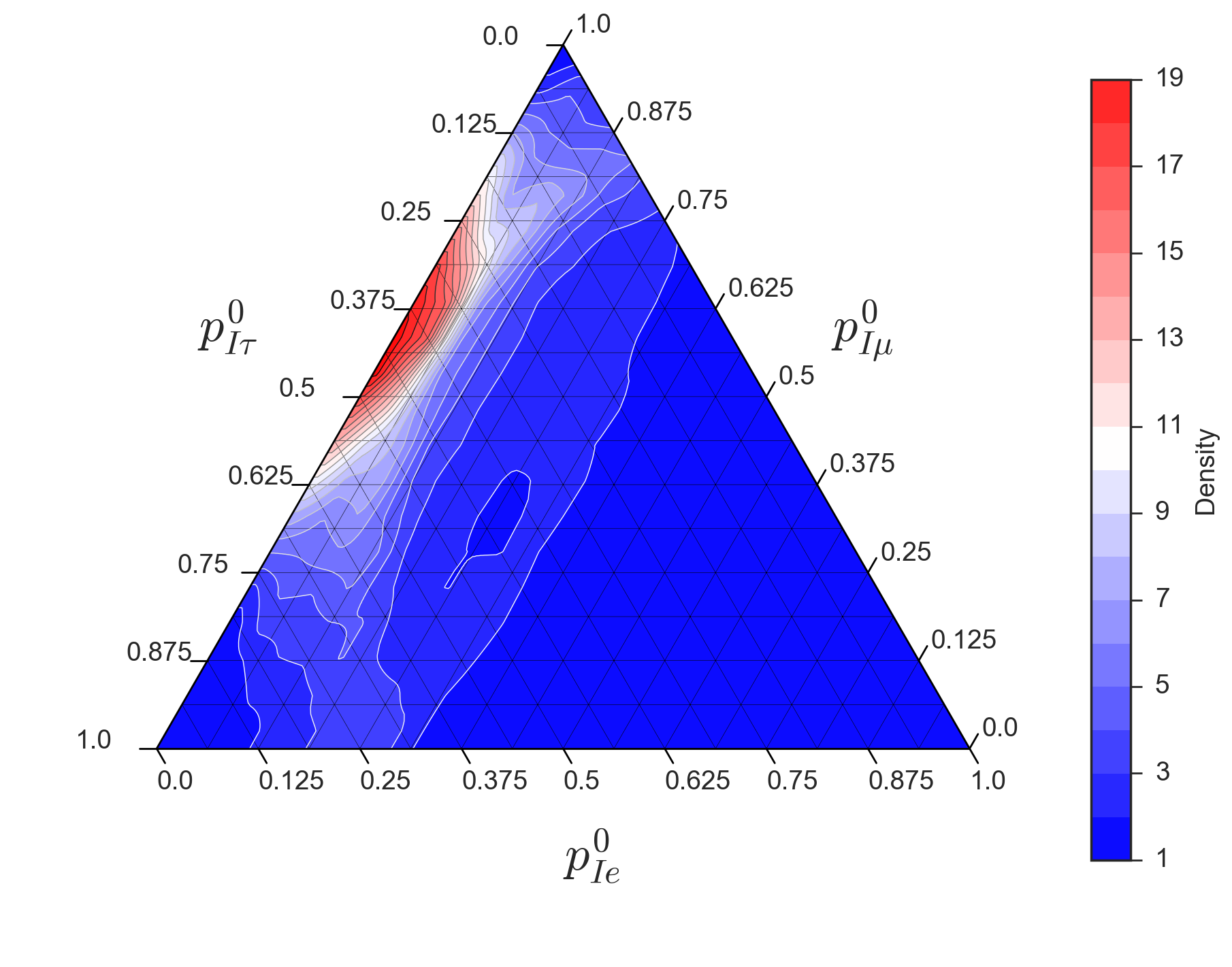}
\vspace{-5mm}
\caption{Density distributions of flavour probabilities $p^0_{{\Romannum{1}}\alpha}$ obtained by a random generation 
of $U$ and $\O$: in the top-left panel mixing angles have been generated randomly uniformly while in the top-right panel
$U$ and $\O$ have been generated according to the Haar measure as explained in the body text. 
In the bottom panels the mixing angles have been generated Gaussianly using the experimental
results still for a uniform generation of $\O$ complex angles
in the left panel or according to the Haar measure in the right panel.}\label{Upmns}
\end{center}
\end{figure}
Some basic results of group theory help explaining why this happens with the usual parameterisation 
and how the problem can be fixed but at the same time they will provide an insightful way to look at 
seesaw models in lepton flavour space. 
 If one looks at the expression (\ref{p0Ial}) in terms of
$U$ and $\O$, then it is clear that the problem is that the usual parameterisations does not 
give a uniform distribution of the elements of $U$ and $\O$.  In order to do that one has
to generate random matrices in a way to cover uniformly the flavour space. 
Let us discuss separately how this can be done for $U$ and $\O$.

\subsection{Random generation of Haar-distributed $U$}

We want to generate $U$ matrices in a way not to privilege any particular lepton flavour basis. 
Let us look again at the panel (a) in Fig.~1.  Here the $U$ matrix is regarded as a (proper) Euler rotation,
an approximate picture that is valid only when phases are neglected.
A flavour unbiased random generation of weak lepton flavour bases, 
has to be such that given a certain flavour axis, for example the tauon axis, this 
points to any infinitesimal surface element  on the unit sphere in lepton flavour space
with equal probability. In this way, generating a large number of
lepton flavour basis, each lepton axis will uniformly cover the unit sphere in lepton flavour space.
This can be done using well known results of group theory that we briefly discuss \cite{rand_uni}.

Each (real) $U$ matrix is an element of the group of real rotations $SO(3,\mathbb{R})$. When phases
are taken into account each randomly generated $U$ is an element of the group of unitary transformations $U(3)$.
Therefore, in group theory language, a flavour unbiased random generation of $U$ corresponds to generate 
randomly unitary matrices according to the Haar measure over the group $U(3)$
({\em Haar-distributed random matrices}) that is  given by
\bea
dV\equiv d(\sin^2\theta_{12})\, d(\sin^2\theta_{23}) \, d(\cos^4\theta_{13}) \, d\delta \,  d\rho \, d\sigma \,  .
\eea
In this way generating uniformly $\sin^2\theta_{12}$,  $\sin^2\theta_{23}$ and 
$\cos^4\theta_{13}$ in the interval $[0,1]$, one obtains equal distributions 
for all $U_{\a i}$ elements, both for their real parts and for their imaginary parts,
as shown in Fig.~3. 
\begin{figure}[H]
\centerline{\includegraphics[scale=0.75]{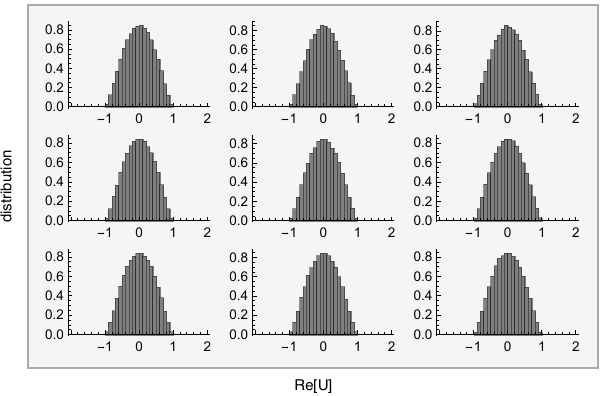} \hspace{5mm}
\includegraphics[scale=0.75]{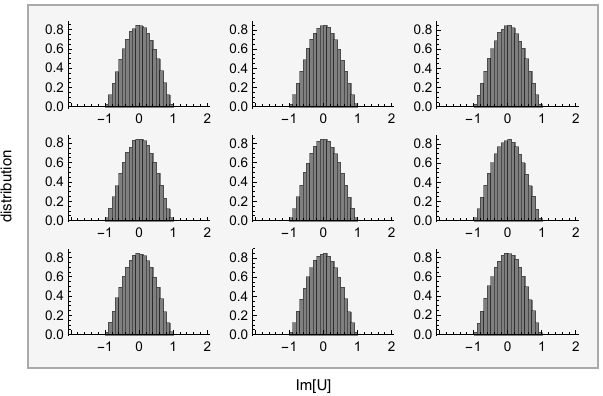}}
\caption{Distributions of  real and imaginary parts of the entries of Haar-distributed $U$.}\label{Upmns}
\end{figure}
Notice that the use of Haar-distributed $U$, and more generally light neutrino mass matrices, is the basis of
anarchical prediction of low energy neutrino parameters \cite{anarchy}. However, for us, more 
pragmatically, this is a way to generate flavoured unbiased scans over seesaw models. 

\subsection{Random generation of Haar-distributed $\O$}

If one uses the experimental information on $U$ and, therefore, the experimental distribution for the  mixing angles,  the random generation of $U$ is not actually necessary, except for the phases that however can be simply generated uniformly between $0$ and $2\pi$ considering that the Haar measure is flat in the phases.

It is then actually more important to generate randomly $\O$'s in a way not to introduce any bias in
flavour space.  Complex orthogonal matrices, as in Eq.~(\ref{Omega}), provide a matrix 
representation of  the Lie group $O(3,\mathbb{C})$ of complex rotations, 
the complex orthogonal group. It is well known that the special group of complex rotations 
$SO(3,\mathbb{C})$, i.e.  those with determinant $+1$,  is isomorphic to the restricted (proper + orthochronous)
Lorentz group  $SO^+(3,1)$.  This can be seen showing that they have the same Lie algebra (see Appendix B).\footnote{Notice that
while the group $U(3)$ of unitary matrices is compact, the group $O(3,\mathbb{C})$ is not, that is 
why they have an intrinsically different parameterisation: in the case of a compact group parameters
always vary within a finite interval, while in the case of a non compact group parameters
can be arbitrarily large. This also leads to an intrinsic different Haar measure for complex orthogonal matrices
compared to unitary matrices.}
 
For this reason a generic complex rotation matrix $\O$ with ${\rm det}(\O)=+1$ can be decomposed as
\bea
\Omega(z_{12},z_{13},z_{23})=R(\a_{12},\a_{13},\a_{23}) \cdot \Omega_{\rm boost}(\vec{\beta}) \,  ,
\eea
where $R$ is a real orthogonal matrix with ${\rm det}(R) = +1$ 
parameterised in terms of three Euler angles $\a_{ij}$ and $\O_{\rm boost}$
is a pure Lorentz boost (in flavour space) parameterised in terms of a 
boost velocity vector $\vec{\beta} =\beta \, \hat{n}$ with an associated Lorentz factor 
$\gamma \equiv (1 -\beta^2)^{-1/2}$. For example, if one chooses a unit vector $\hat{n} = (0,0,1)$, 
then one simply has\footnote{See Appendix B for details.}
\bea\label{boostthirdaxis}
\Omega_{\rm boost}(0,0,\b)=\begin{pmatrix}
\cosh\psi & -i \sinh\psi &0 \\ i \sinh\psi & \cosh\psi &0\\ 0&0& 1
\end{pmatrix}  \, ,
\eea
with $\beta=\tanh \psi$ and $\g = \cosh\psi$.  
This special case can be of course generalised for an arbitrary choice of $\hat{n}$ (see Appendix B).
It is interesting to notice that for transformations with
$\b \neq 0$ there is a privileged direction in flavour space while transformations with $\beta =0$ corresponds
basically to  a flavour invariant situation where $\O = B = P$ and the fine-tuning in the seesaw is minimum.\footnote{We are barring 
the real rotation component $R(\a_{12},\a_{13},\a_{23})$.}
Indeed, notice that the see-saw  fine-tuning parameters associated to the light neutrino masses
introduced in Section 2 are in this case simply given by $\gamma_1 = \gamma_2 =\gamma^2\,(1+\b^2)$, 
showing that  the Lorenz factor of the transformation is related to the fine-tuning parameters.
This is somehow  another way to understand why imposing a flavour symmetry leads to $\O=B = P$: this is
the case corresponding to vanishing velocity in flavour space, meaning that the model
does not have any privileged flavour direction.

In the case of the special boost in Eq.~(\ref{boostthirdaxis}), the bridging matrix is given by
\be
B_{\rm boost} = 
\begin{pmatrix}
{\sqrt{m_1}\,\cosh\psi \over \sqrt{m_1\,\cosh^2 \psi + m_2 \, \sinh^2 \psi}} & 
{i\,\sqrt{m_1}\,\sinh\psi \over  \sqrt{m_1\,\sinh^2 \psi + m_2 \, \cosh^2 \psi}} & 0 \\
{i\,\sqrt{m_2}\,\sinh\psi \over \sqrt{m_1\,\sinh^2 \psi + m_2 \, \cosh^2 \psi}} & 
{\sqrt{m_2}\,\cosh\psi \over \sqrt{m_1\,\sinh^2 \psi + m_2 \, \cosh^2 \psi}} &0 \\ 
0&0& 1
\end{pmatrix}  \,  ,
\ee
an example confirming that the heavy lepton flavour basis is in general non-orthonormal.

If in the orthogonal matrix we turn on, in addition to a boost, a real rotation,
then in the limit $\beta = 0$ one  obtains $\O= R(\a_{12},\a_{13},\a_{23})$, i.e., in general one  
does not recover form dominance corresponding to $\O = P$. 
For example, let us consider  a simple rotation around the 
third axis ($\a_{13}=\a_{23}=0$), so that simply
\be
\O = \begin{pmatrix}
\cos\a_{12} & \sin\a_{12} &0 \\ -\sin\a_{12} & \cos \a_{12} &0\\ 0&0& 1
\end{pmatrix}  \,  .
\ee
This case still corresponds to  a case of minimal fine tuning, since one clearly 
has $\gamma_1 = \gamma_2 = \gamma_3 = 1$. However, in this case  one finds for the bridging matrix
\be
B = 
\begin{pmatrix}
{\sqrt{m_1}\,\cos\a_{12}\over \sqrt{m_1\,\cos^2 \a_{12} + m_2 \, \sin^2 \a_{12} }}& 
{\sqrt{m_1}\,\sin\a_{12}\over \sqrt{m_1\,\sin^2 \a_{12} + m_2 \, \cos^2 \a_{12} }} &0 \\
-{\sqrt{m_2}\,\sin\a_{12}\over \sqrt{m_1\,\cos^2 \a_{12} + m_2 \, \sin^2 \a_{12} }}  & 
{\sqrt{m_2}\,\cos\a_{12}\over \sqrt{m_1\,\sin^2 \a_{12} + m_2 \, \cos^2 \a_{12} }}  &0 \\ 0&0& 1
\end{pmatrix}  \,  , 
\ee 
showing that this, in general, does not coincide with the orthogonal matrix (it is not a real
rotation matrix) and also that it  brings to an heavy lepton flavour basis that is not orthonormal. 
These kind of models, with a real orthogonal matrix coinciding with a rotation matrix, are interesting since they still correspond 
to minimal fine-tuning but, for the basis is not orthonormal, there can be in general interference among heavy neutrino flavours so that the flavoured
 $C\!P$ asymmetries do not vanish in general. In this way in principle one could have leptogenesis stemming entirely from low energy 
neutrino phases \cite{realomega}.
However, unless one has a strong resonance enhancement, the observed asymmetry is usually not reproduced, 
implying that one needs to turn a boost on as well. Therefore, it seems that the explanation of the matter-antimatter asymmetry of the universe  
necessarily requires the existence of some privileged direction in lepton flavour space, corresponding to
some mismatch between the bases where the Majorana  and the Yukawa mass matrices are diagonal.
 
If we want again to generate flavour unbiased $\O$ matrices, it is then clear what we have to do now.
First of all one has to generate Haar-distributed rotation matrices $R(\a_{12},\a_{13},\a_{23})$ 
as we did for $U$.  For $SO(3,\mathbb{R})$,  the Haar measure is quite simple
\bea
dV\equiv d(\sin \a_{13}) \, d\a_{23} \, d\a_{12}.
\eea
In the case of  $\Omega_{\rm boost}(\vec{\b})$, it is clear that, for a fixed value of $\beta$, 
we need to generate isotropically unit vectors $\hat{n}$. For example, one can use polar coordinates and write 
\be
\hat{n}=(\cos\phi\sin\theta,~\sin\phi\sin\theta,~\cos\theta),
\ee
with $\theta\in[0,\pi]$ and $\phi\in[0,2\pi]$. If one generates random uniformly $\theta$ and $\phi$, one would obtain
a clustered density of points near the poles. To generate isotropic orientation of $\hat{n}$, we use the standard technique of generating uniform points on a surface of a sphere by generating values of $\theta$ given by
\be
\theta=\cos^{-1}(1-2\,a)  \,  ,
\ee
with the parameter $a$ uniformly generated within the interval $[0,1]$.
Using this random generation procedure we have obtained
distributions for the real and imaginary parts of the orthogonal matrix plotted in Fig.~4
\begin{figure}[H]
\centerline{
\includegraphics[scale=0.75]{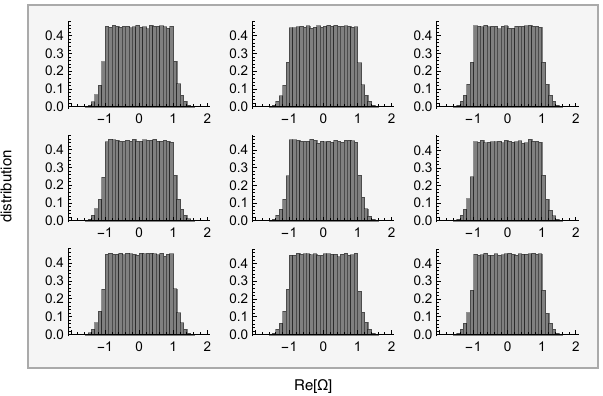} \hspace{5mm}
\includegraphics[scale=0.75]{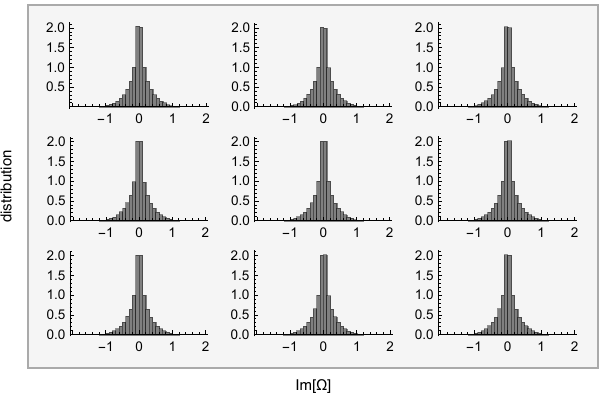}}
\caption{Distribution of the real and imaginary parts of the entries $\Omega_{iJ}$.}\label{omg}
\end{figure}
Notice that this procedure can be easily extended to include also matrices $\O$ with negative determinant.

Finally, we have combined together the flavour unbiased procedure to random generate both $U$ and $\O$
and again plotted the distribution of probabilities $p^0_{I\a}$ shown in the top right panel of Fig.~2 and, as one can see, this time,
barring small statistical fluctuations, we have obtained a perfectly flavour homogeneous distribution of points shown in
the top-right panel. 

In the bottom panels of Fig,~2 we plotted the $p^0_{I\a}$ using the experimental information, 
generating the mixing angles
in $U$ random Gaussianly  and using the following latest experimental results for the values of the 
mixing angles in the case of normal ordering \cite{nufit}
\bea \label{exp}
\theta_{12}  & = & 33.82^{\circ} \pm 0.77^{\circ} \,  ,\\ \nonumber
\theta_{13}  & = &   8.61^\circ \pm 0.12^\circ \,  , \\  \nonumber
\theta_{23} & = &   49.7^\circ \pm 1.0^\circ \,   .
\eea
The inverted ordering case is now disfavoured at more than $3\,\s$ and we will not consider it in our
following discussion.  We have also compared again the case when mixing angles are random uniformly generated (left panel)  with the case of Haar-distributed $U$ (right panel).
This time one can see that there is not a great difference since in any case the region 
that is biased is disfavoured by current data.  
 
 \subsection{An application: $N_2$-leptogenesis}
 
The reason why we focused in Fig.~2 on $p^0_{{I}\a}$, is that the lightest RH neutrino plays a 
particular role in $N_2$-leptogenesis \cite{geometry,vives}. In this scenario of leptogenesis 
the current baryon asymmetry, expressed in terms of the baryon-to-photon number ratio at present $\eta_{B 0}$, 
can be calculated as \cite{riotto1,riotto2}
\bea
\eta_{B 0} & \simeq & 0.96 \times 10^{-2}\, 
\left(\ve_{{\Romannum{2}}e} \, \kappa(K_{{\Romannum{2}}e} +
K_{{\Romannum{2}}\m}) \, e^{-{3\pi\over 8}\,K_{{I} e}}+ \right. \\
& + & \left. \ve_{{\Romannum{2}}\mu} \, \kappa(K_{{\Romannum{2}}e} + 
K_{{\Romannum{2}}\m}) \, e^{-{3\pi\over 8}\,K_{{I} \mu}}
+ \ve_{{\Romannum{2}}\tau} \, \kappa(K_{{\Romannum{2}}\tau}) \, 
e^{-{3\pi\over 8}\,K_{{I} \tau}}
\right)\,  .
\eea
The lightest RH neutrino flavoured decay parameters $K_{{I} \a}$ play clearly
a special role since they describe the exponential wash-out from lightest RH neutrino inverse decays
and one needs that at least one of them is less than unity for
the asymmetry produced by the $N_2$-decays at a temperature $T\sim M_2$ to survive at present. 
The flavoured decay parameters are simply given by $K_{I\a} = p^0_{I\a} \, K_I$,
where $K_I = \widetilde{m}_I/m_{\star}$ are the total decay parameters and
$m_{\star}$ is the equilibrium neutrino mass.  Therefore, one can see the 
special role played by the $p^0_{{I}\a}$'s in $N_2$-leptogenesis. 

It is then particularly interesting to understand how special is the condition for the asymmetry produced
by $N_2$ decays to survive at present. This is basically equivalent to understand how special is to 
have at least one $K_{{I} \a}$, for some lepton flavour $\a$, less than unity. 

The flavoured decay parameters are related to the orthogonal matrix 
through the Eq.~(\ref{effectivenmasses}) for $\widetilde{m}_I$ and, therefore, for each 
choice of $\O$ and for a given value of $m_1$, one has a corresponding set of values of $K_{{I}\a}$.
We have therefore produced the distributions for the values of the $K_{{I}\a}$ for $\a=e,\mu,\tau$
adopting the flavoured unbiased procedure, based on the Haar measure, that we discussed. 
In Fig.~5 the distributions are shown without imposing any experimental information 
on the values of the mixing angles that, therefore, vary arbitrarily within  
$[0,90^{\circ}]$ and in the hierarchical limit $m_1 =0$.
It can be seen how the distributions are identical independently of $\a$ as a result of
the flavour blindness of the  procedure we followed to generate randomly $U$ and $\O$.\footnote{We do not
show the distributions for the $K_{{\Romannum{2}}\a}$ 
and for the $K_{{\Romannum{3}}\a}$ but they would also be identical since the procedure is 
flavour blind both to charged lepton flavour and to heavy neutrino flavour.}
\begin{figure}
\includegraphics[scale=0.4]{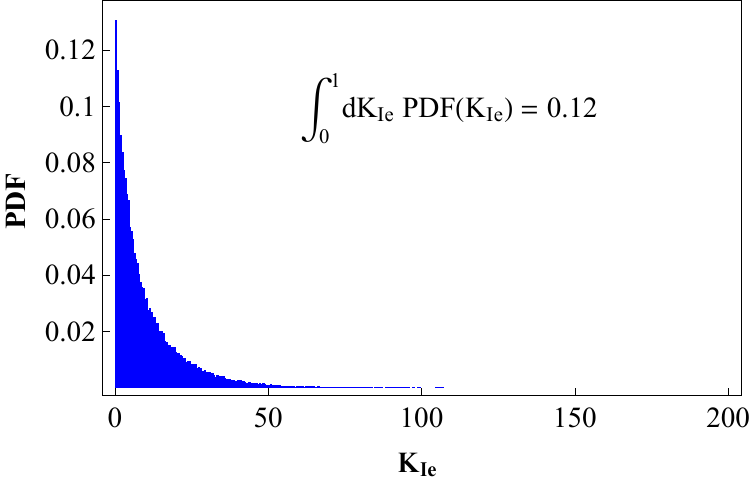}
\includegraphics[scale=0.4]{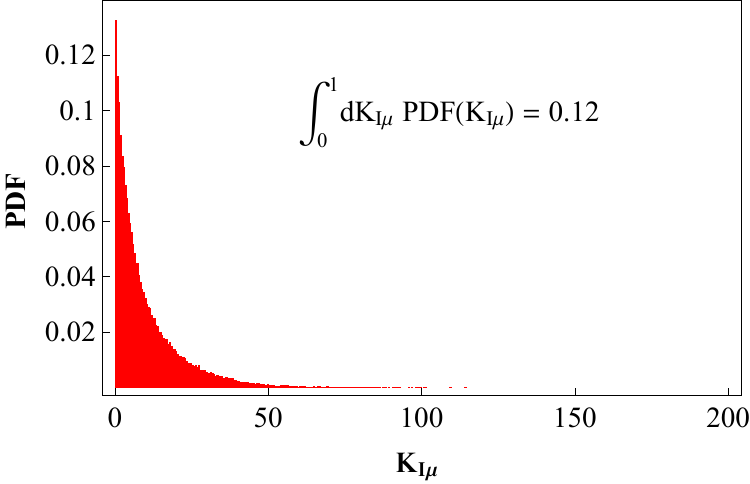}
\includegraphics[scale=0.4]{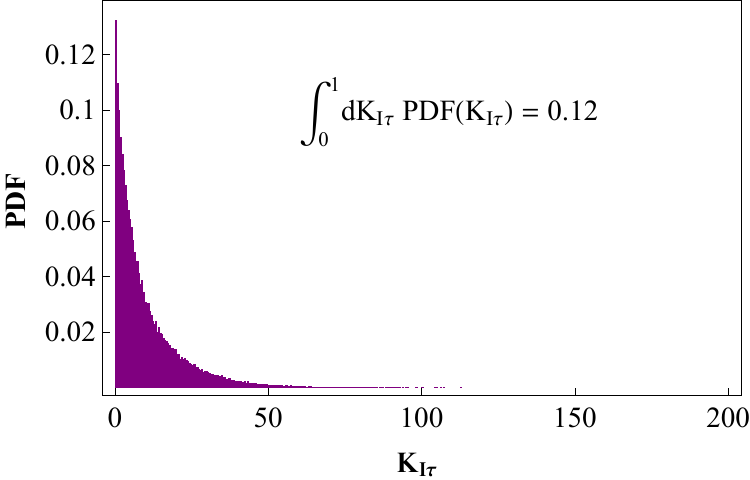}
\label{dist_decayA}
\caption{Distributions of the decay parameters for $m_1 =0$ and generic values of the mixing 
from a  random flavour blind generation of $U$ and $\O$.}
\end{figure}
It is important to notice that the probability for each $K_{{I}\a}$ to be less than unity
is about $12\%$, meaning that the probability that at least one  $K_{{I}\a}$
is less than unity is approximately $36\%$.  

How do these results change when the experimental information on the mixing angles is used?
In Fig.~6 the distributions for the $K_{{I}\a}$ are now obtained using the experimental information
on the mixing angles Eq.~(\ref{exp}). One can see how the fact that the experimental
values favour small values of $p^0_{I\a}$, translates into a much higher probability, approximately $36\%$, 
for $K_{{I}e}$ to be less than unity compared to $K_{{I}\mu}$ and $K_{{I}\tau}$
whose probability to be less than unity drops to $\sim 6$--$7\%$. 
\begin{figure}
\includegraphics[scale=0.4]{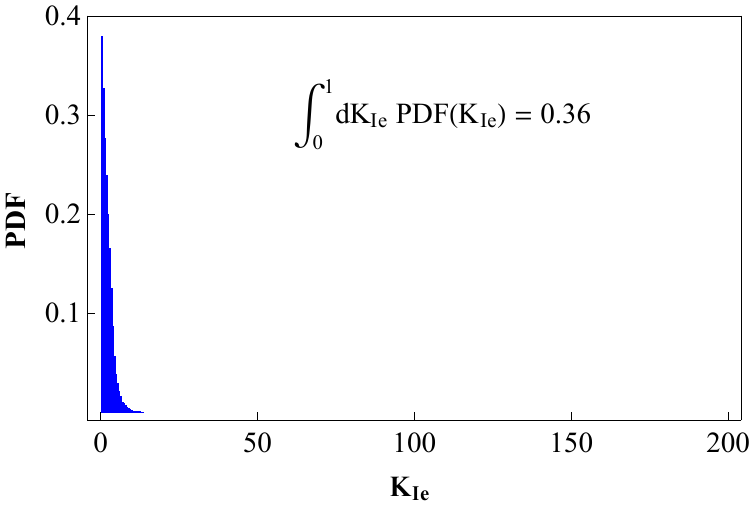}
\includegraphics[scale=0.4]{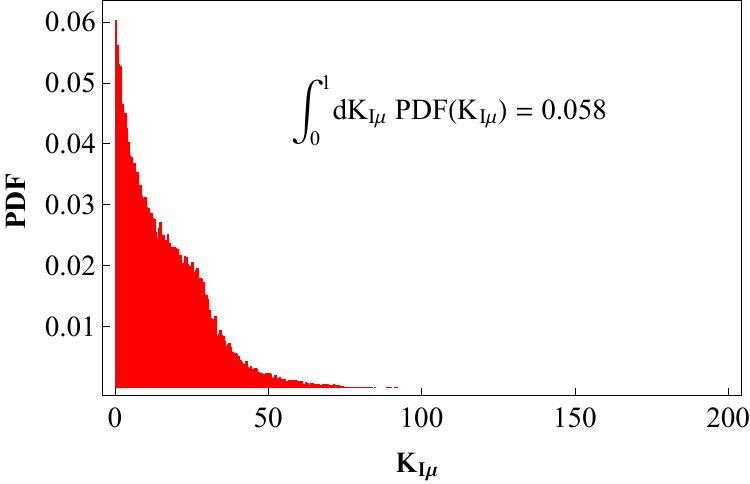}
\includegraphics[scale=0.4]{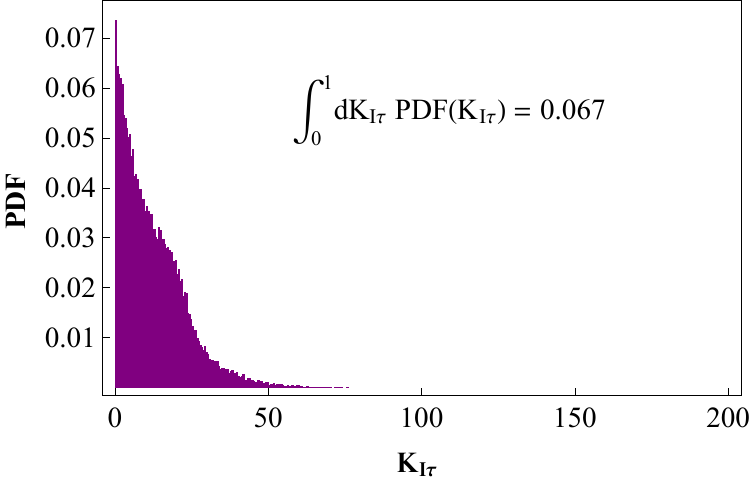}
\caption{Same as in Fig.~5 but using the experimental values of
the mixing angles in Eq.~(\ref{exp}).}\label{dist_decayB}
\end{figure}
The probability that at least one $K_{{I}\a}$ is less than unity is therefore now about $49\%$.
This result shows how the possibility to have a small lightest RH neutrino  wash-out in one of the three flavours,
a crucial condition to realise $N_2$-leptogenesis, is not special at all (contrarily to some statements made in the literature). 
It should be also said that on the other hand the probability to have
$K_{I} = \sum_\a \, K_{{I}\a} < 1$ is only $0.1 \%$ confirming and quantifying 
how accounting for flavour effects is crucial for $N_2$-leptogenesis \cite{vives}. 

In Fig.~7 and Fig.~8, again for arbitrary and experimental values of the mixing angles respectively, 
we also show how the distributions change departing from the hierarchical limit, for
$m_1 =0.01\,{\rm eV}$.  
\begin{figure}
\includegraphics[scale=0.4]{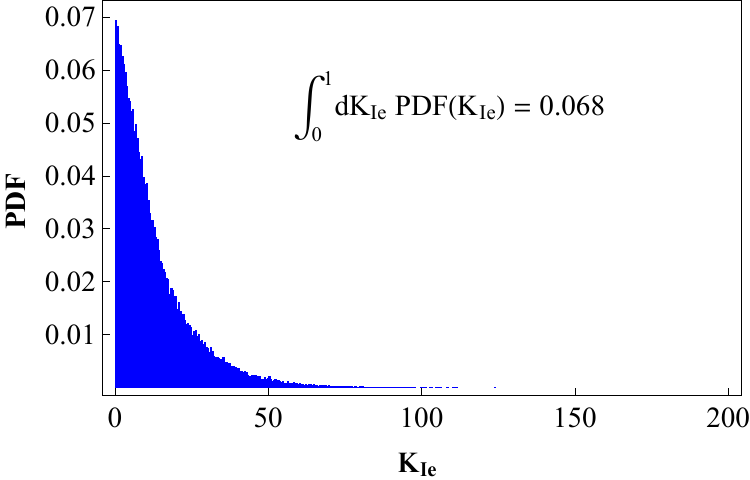}
\includegraphics[scale=0.4]{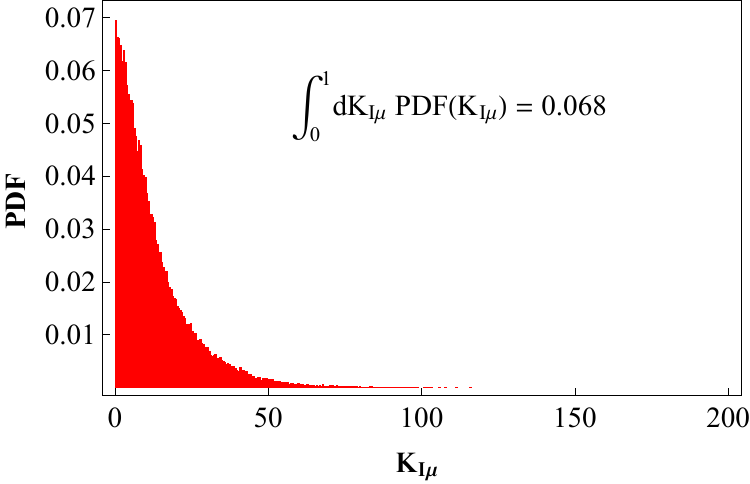}
\includegraphics[scale=0.4]{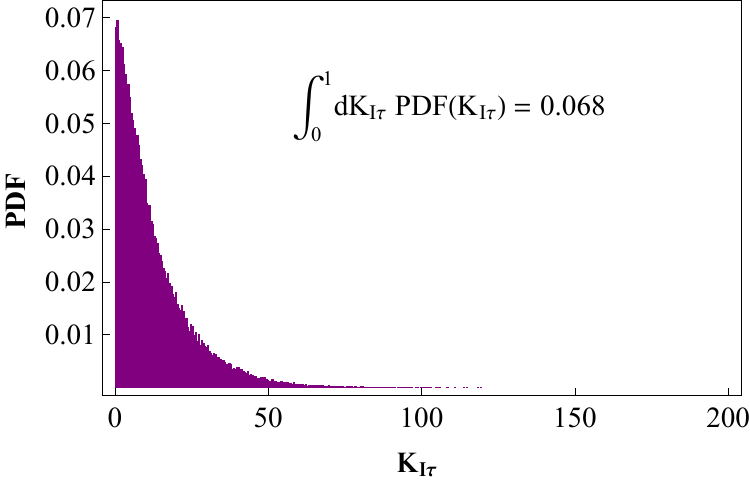}
\caption{Same as in Fig.~5 but for $m_1 =0.01\,{\rm eV}$.}\label{dist_decayC}
\end{figure}
It can be noticed how all probabilities drop and this is easily explained since $K_{I} = 
\widetilde{m}_I /m_{\star}$ and $\widetilde{m}_{I} \geq m_1$ so that all decay parameters
tend to increase.  However, the probability that at least one of the three $K_{{I}\a} < 1$
is still quite large, approximately $23\%$.
Notice also how for arbitrary mixing angles the distributions are still identical in the
three flavours.
\begin{figure}
\includegraphics[scale=0.4]{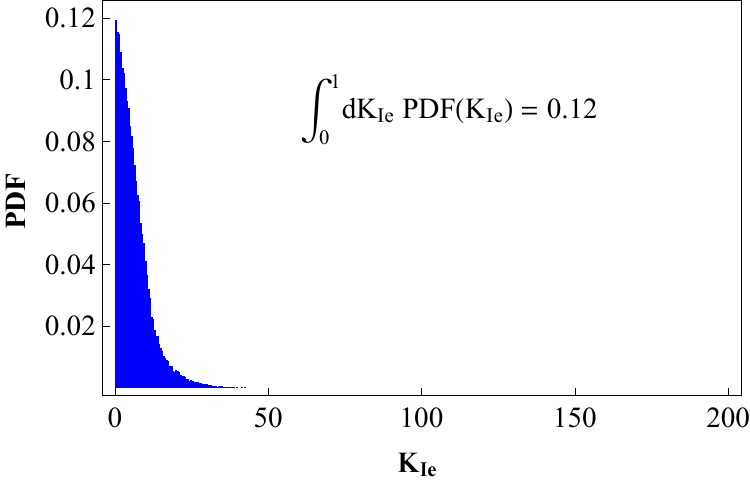}
\includegraphics[scale=0.4]{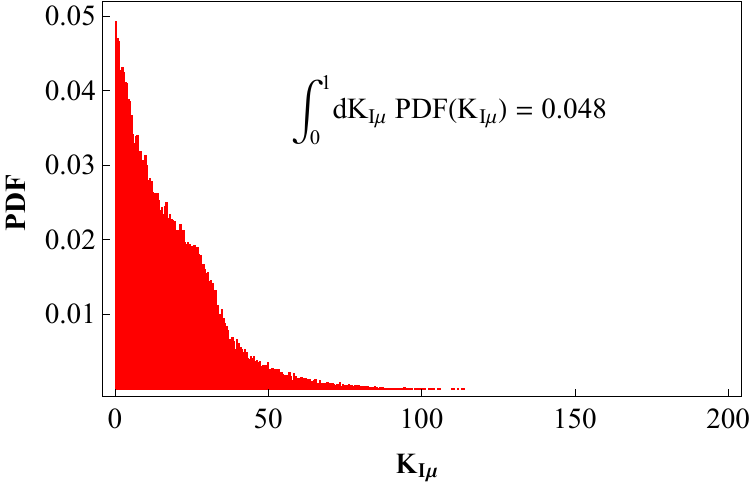}
\includegraphics[scale=0.4]{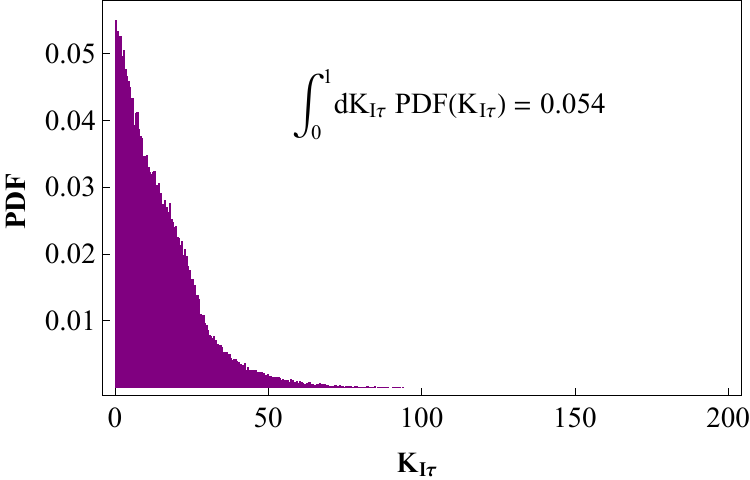}
\caption{Same as in Fig.~6 but for $m_1 =0.01\,{\rm eV}$.}\label{dist_decayD}
\end{figure}
For values $m_1 \gtrsim 0.1\,{\rm eV}$, the probability that at least one  $K_{{I}\a} <1$ drops below $5\%$. 
This can somehow be regarded as a kind
of extension of the upper bound on neutrino masses holding in $N_1$ leptogenesis, also
to the case of $N_2$-leptogenesis, though it should be clear that in this case 
the upper bound should be interpreted more in a statistical
way rather than as an absolute one. 

\section{Conclusion}

We have seen how representing seesaw models in lepton flavour space allows a deeper understanding  of different
features of seesaw models. In particular different results have an easy graphical interpretation. The new
bridging matrix, switching from the light to heavy neutrino lepton basis, is very useful for writing flavour probabilities  and characterising seesaw models in an easy way. We have also seen how fine-tuning in seesaw models can be expressed in terms of a Lorentz boost in flavour space and sequential dominated models, characterised by minimal fine-tuning, are those for which the boost velocity is vanishing and therefore are flavour invariant, in agreement with previous results. A deviation from sequential dominated models, turning on some motion in flavour space,
produces a deviation from orthonormality that is necessary to have non-vanishing $C\!P$ decaying asymmetries
and successful  leptogenesis.  We have seen also that one can deviate from sequential dominated models
with a pure real rotation orthogonal matrix. This still corresponds to models at rest in flavour space, but 
again with some deviation from orthonormality
producing non-vanishing $C\!P$ violation and in principle allowing for successful leptogenesis.  We have also seen how to generate
randomly, in a flavour unbiased way, seesaw models. We have then applied these new tools to $N_2$-leptogenesis
and showed how it is actually very easy to realise the condition of no wash-out from the lightest RH neutrino,
for $\sim 49\%$ of the points once the current values of the mixing angles are used for 
hierarchical light neutrinos. On the other hand, if $m_1 \gtrsim 0.1\,{\rm eV}$, this probability drops 
to less than $5\%$, a result that confirms that the exclusion of quasi-degenerate neutrino from
current cosmological observations,  supports scenarios of minimal leptogenesis based on type-I seesaw and thermal production of RH neutrinos.  The new tools and ideas introduced in this work will be
very useful in different respects, both in the quest of models of new physics able to explain neutrino masses and mixing and, more pragmatically, in scanning seesaw models within different contexts such as leptogenesis. 

\vspace{-1mm}
\subsection*{Acknowledgments}

We wish to thank Steve King for pointing out useful references. 
PDB  acknowledges financial support from the STFC Consolidated Grant L000296/1. 
RS is supported  by a Newton International Fellowship (NF 171202) from Royal Society (UK) and SERB (India). 
This project has received funding/support from the European Union Horizon 2020 research and innovation 
programme under the Marie Sk\l{}odowska-Curie grant agreements number 690575 and  674896.
This work was performed in part at Aspen Center for Physics, which is supported by National Science Foundation grant PHY-1607611.
 
\section*{Appendix A}
\appendix
\renewcommand{\thesection}{\Alph{section}}
\renewcommand{\thesubsection}{\Alph{section}.\arabic{subsection}}
\def\theequation{\Alph{section}.\arabic{equation}}
\renewcommand{\thetable}{\arabic{table}}
\renewcommand{\thefigure}{\arabic{figure}}
\setcounter{section}{1}
\setcounter{equation}{0}

In this Appendix we list explicitly the forms of $m_D$ corresponding to 
the nine cases discussed in Section 2 with two heavy neutrino flavours 
coinciding with a charged lepton flavour, so that 
$|L_I \rangle=|L_J \rangle = |L_\alpha\rangle$ with
$I \neq J$ and $\a =e,\mu,\tau$. These are given by

{\small 
\bea \nonumber
m_D   & = & 
\left( \begin{array}{ccc}
m_{De {\Romannum{1}}} & m_{De \Romannum{2}}  & m_{De {\Romannum{3}}}   \\
0 & 0 & m_{D \mu  {\Romannum{3}}}  \\
0 & 0 & m_{D \tau  {\Romannum{3}}} 
\end{array}\right)  \,  , \;\;
\left( \begin{array}{ccc}
m_{De {\Romannum{1}}} & m_{De \Romannum{2}}  & m_{De {\Romannum{3}}}   \\
0 & m_{D \mu  {\Romannum{2}}}  &  0 \\
0 & m_{D \tau  {\Romannum{2}}}  &  0
\end{array}\right)  \,  , \;\;
\left( \begin{array}{ccc}
m_{De {\Romannum{1}}} & m_{De \Romannum{2}}  & m_{De {\Romannum{3}}}   \\
m_{D \mu  {\Romannum{1}}}  & 0  &  0 \\
m_{D \tau  {\Romannum{1}}} & 0  &  0
\end{array}\right) \,  ,   \\ \nonumber
&  &
\left( \begin{array}{ccc}
0  & 0 &  m_{D \tau  {\Romannum{1}}} \\
m_{D\mu {\Romannum{1}}} &  m_{D\mu \Romannum{2}}   &  m_{D\mu {\Romannum{3}}}  \\
0 & 0 & m_{D \tau  {\Romannum{3}}} 
\end{array}\right)  \,  , \;\;
\left( \begin{array}{ccc}
0 & m_{D \mu  {\Romannum{2}}}  &  0 \\
m_{D\mu {\Romannum{1}}} & m_{D\mu \Romannum{2}}  & m_{D\mu {\Romannum{3}}}   \\
0 & m_{D \tau  {\Romannum{2}}}  &  0
\end{array}\right)  \,  , \;\;
\left( \begin{array}{ccc}
m_{D e  {\Romannum{1}}}  & 0  &  0 \\
m_{D\mu {\Romannum{1}}} & m_{D\mu \Romannum{2}}  & m_{D\mu {\Romannum{3}}}   \\
m_{D \tau  {\Romannum{1}}} & 0  &  0
\end{array}\right) \,  ,   \\ \nonumber
&  &
\left( \begin{array}{ccc}
0  & 0 &  m_{D e  {\Romannum{3}}} \\ 
0 & 0 & m_{D \mu {\Romannum{3}}}  \\
m_{D\tau {\Romannum{1}}} &  m_{D\tau \Romannum{2}}   &  m_{D\mu {\Romannum{3}}}  \\
\end{array}\right)  \,  , \;\;
\left( \begin{array}{ccc}
0 & m_{D e  {\Romannum{2}}}  &  0 \\
0 & m_{D \mu  {\Romannum{2}}}  &  0 \\
m_{D\tau {\Romannum{1}}} & m_{D\tau \Romannum{2}}  & m_{D\tau {\Romannum{3}}}   \\
\end{array}\right)  \,  , \;\;
\left( \begin{array}{ccc}
m_{D e {\Romannum{1}}}  & 0  &  0 \\
m_{D \mu  {\Romannum{1}}} & 0  &  0 \\
m_{D\tau {\Romannum{1}}} & m_{D\tau \Romannum{2}}  & m_{D\tau {\Romannum{3}}}   \\
\end{array}\right) \,  .  \\
\eea
}
All these cases are excluded, since
they give rise to a light neutrino mass matrix that is either again respecting the scaling ansatz made in \cite{moharode} or has some similar scaling property also leading to unacceptable low energy neutrino data.
 Let us give here a few more details.  If, for example, $\a=e$, then 
the resulting light neutrino Majorana mass matrix ($m_\nu$) is of a special form obeying
{\em strong scaling ansatz} that is ruled out\cite{moharode}.  Basically this corresponds to a situation
when in the light neutrino Majorana matrix  one of the rows  is $c$ times of an other row, where, `$c$' is a common scale factor which can be expressed as a function of the elements of $m_D$. For example, if
$I=\Romannum{1}$ and  $J=\Romannum{2}$, then  the second row  is  $c$ times the third row. This  leads to a vanishing eigenvalue and the corresponding eigenvector has one vanishing entry. This results  in vanishing $U_{e3}$ or $U_{e1}$, depending 
whether one has inverted or the normal mass ordering. It can be checked that the other two cases, 
i.e., $(I,J)=(\Romannum{1},\Romannum{3})$ and $(I,J)=(\Romannum{2},\Romannum{3})$, 
also  lead to a form obeying the strong scaling ansatz (though with different scale factors) and, 
as in the previous case, this results into vanishing $U_{e3}$ or $U_{e1}$. 
In the the  remaining six cases, for $\a =\mu,\tau$, we found analogously that either $U_{\alpha 3}$ or $U_{\alpha 1}$ vanishes, again for inverted and normal mass ordering respectively. 
Since a zero entry in $U$ is  excluded by the experimental data,  
we conclude that all nine cases corresponding to have two coinciding heavy neutrino flavours are ruled out. 

\section*{Appendix B}
\appendix
\renewcommand{\thesection}{\Alph{section}}
\renewcommand{\thesubsection}{\Alph{section}.\arabic{subsection}}
\def\theequation{\Alph{section}.\arabic{equation}}
\renewcommand{\thetable}{\arabic{table}}
\renewcommand{\thefigure}{\arabic{figure}}
\setcounter{section}{1}
\setcounter{equation}{0}

In this Appendix we generalise the parameterisation of the orthogonal matrix in terms of 
three real angles and the three components of the boost velocity. 
As well known, a generic proper orthochronous Lorentz transformation  can be written
as $\L = e^{-i\,(\vec{\a}\cdot \vec{J} + \vec{\xi}\cdot \vec{K}})$ where $\vec{J}$ and $\vec{K}$ 
are respectively the rotation and boost generators of $SO^+(3,1)$ 
and obey the Lie algebra\footnote{Both $\vec{J}$ and $\vec{K}$ are represented
by $4 \times 4$ matrices.}
\begin{eqnarray}
\left[J_i, J_j \right]  & = &  i \, \epsilon_{ijk} \, J_k \,  ,  \\
\left[K_{i}, K_j \right]  & =  & -i \, \epsilon_{ijk} \, K_k  \,  ,  \\
\left[J_i, K_j \right]  & = & i\, \epsilon_{ijk}   \,  K_k  \,  .
\end{eqnarray}
where $\epsilon_{ijk}$ is the totally antisymmetric tensor. 
If we now consider a generic complex rotation $\Omega$ belonging to $SO(3,\mathbb{C})$, 
this can be written as $\Omega=e^{-i\,(\vec{\a}\cdot \vec{L}+\vec{\xi}\cdot \vec{\S}})$, where $\vec{L}$
are the generators of the real rotations and $\vec{\Sigma} = i \, \vec{L}$ are the generators
of the imaginary rotations (hyperbolic rotations), both represented by $3 \times 3$ matrices.  
The generators of $SO(3,\mathbb{C})$ satisfy the Lie algebra
\bea
\left[L_i, L_j \right]=i \, \epsilon_{ijk}\,L_k \,  , \\
\left[\Sigma_i,\Sigma_j \right]=-i\epsilon_{ijk} \, \Sigma_k \, , \\
\left[L_i, \Sigma_j \right]= i \, \epsilon_{ijk}\Sigma_k  \,  ,
\eea
clearly coinciding with that one of  $SO(3,1)^+$, with the identification $\vec{J} \leftrightarrow \vec{L}$ and 
$\vec{K} \leftrightarrow \vec{\Sigma}$. Therefore,  $\vec{\Sigma}$ correspond to the boost generators
in $SO(3, \mathbb{C})$. 
This shows that the two groups are indeed isomorphic and one can find a map between each other. 
We can then decompose the $3\times 3$ complex 
orthogonal matrix $\Omega \in SO(3,\mathbb{C})$ as
\bea
\Omega=R(\vec{\a}) \cdot \Omega_{\rm boost}(\vec{\beta}),
\eea
where the $3 \times 3$ matrices  $R(\vec{\a}) = e^{-i\,\vec{\a}\cdot \vec{L}} \in SO(3,\mathbb{R})$ 
and $\Omega_{\rm boost} = e^{-i\,\vec{\xi}\, \cdot \, \vec{\S}}$ are generated by 
$\vec{L}$ and $\vec{\Sigma}$ respectively and we defined $\vec{\a}\equiv (\a_{12},\a_{13},\a_{23})$. 
Since $\vec{\S} = i\,\vec{L}$, then  one has simply 
$\O_{\rm boost}(\vec{\b}) = e^{-i\,(i\,\vec{\xi})\cdot \vec{L}}$, showing that boosts are
rotations with complex angles $i\,\vec{\xi}$. In the special case $\vec{\beta}=(0,0,\b)$, considered in the body text,
one obtains easily Eq.~(\ref{boostthirdaxis}). In general, 
for $\vec{\beta}=\beta\,\hat{n}$ and $\hat{n}=(n_1, n_2, n_3)$, one obtains
{\tiny
\bea
\Omega_{\rm boost}(\vec{\beta})=\begin{pmatrix}
\cosh\xi+n_1^2(1-\cosh\xi)&n_1 \, n_2 \, (1-\cosh\xi)-i \, n_3 \sinh\xi & n_1 \, n_3 (1-\cosh\xi)+i \, n_2 \sinh\xi \\ 
n_1 \, n_2 \, (1-\cosh\xi)+i \, n_3 \sinh\xi & \cosh\xi+n_2^2(1-\cosh\xi) & n_2 \, n_3 \, (1-\cosh\xi)-i \, n_1 \, \sinh\xi \\ 
n_1 \, n_3 \, (1-\cosh\xi)-i \, n_2 \sinh\xi & n_2 \, n_3 \, (1-\cosh\xi)+i \, n_1 \, \sinh\xi & \cosh\xi+n_3^2(1-\cosh\xi)  \,  ,
\end{pmatrix},\label{boost2}
\eea
}
with $\b = \tanh \xi$.

\end{document}